\documentclass[fleqn,usenatbib]{mnras}

\usepackage{newtxtext,newtxmath}
\usepackage[T1]{fontenc}
\usepackage{ae,aecompl}
\usepackage[switch,columnwise]{lineno}
\usepackage{bm}
\usepackage[dvipdfmx]{graphicx}	
\usepackage{amsmath}	
\usepackage{amssymb}	

\title[Two-temperature MHD simulation for AGN jets]
{Two-temperature Magnetohydrodynamic simulations for sub--relativistic AGN jets: \\
 Dependence on the fraction of the electron heating }

\author[T. Ohmura et al.]{
Takumi Ohmura,$^{1}$\thanks{E-mail: ohmura@phys.kyushu-u.ac.jp}
Mami Machida,$^{2}$
Kenji Nakamura,$^{3}$
Yuki Kudoh, $^{4}$
\newauthor
~~and Ryoji Matsumoto$^{5}$
\\
$^{1}$Graduate School of Science, Kyushu University, 744 Motooka, Nishi-ku, Fukuoka, Fukuoka 819-0395, Japan\\
$^{2}$Department of Physics, Faculty of Sciences, Kyushu University, 744 Motooka, Nishi-ku, Fukuoka 819-0395, Japan\\
$^{3}$Department of Mechanics, Faculty of Science and Technology, Kyushu Sangyo University,\\ 3-1 Matsukadai 2-chome, Higashi-ku, Fukuoka 813-8503, Japan\\
$^{4}$Graduate School of Science and Engineering, Kagoshima University, Kagoshima 890-0065, Japan\\
$^{5}$Department of Physics, Graduate School of Science, Chiba University, 1-33 Yayoi-cho, Inage-ku, Chiba, 263-8522, Japan
}

\date{Accepted XXX. Received YYY; in original form ZZZ}

\pubyear{2020}

\begin{document}
\label{firstpage}
\pagerange{\pageref{firstpage}--\pageref{lastpage}}
\maketitle

\begin{abstract}
We present the results of two-temperature magnetohydrodynamic simulations of the propagation of sub-relativistic jets of active galactic nuclei.
The dependence of the electron and ion temperature distributions on the fraction of electron heating $f_{\rm e}$ at the shock front is studied for $f_{\rm e}=0, 0.05$, and 0.2.
Numerical results indicate that in sub-relativistic, rarefied jets, the jet plasma crossing the terminal shock forms a hot, two-temperature plasma in which the ion temperature is higher than the electron temperature.
The two-temperature plasma expands and forms a backflow referred to as a cocoon, in which the ion temperature remains higher than the electron temperature for longer than 100 Myr.
Electrons in the cocoon are continuously heated by ions through Coulomb collisions, and the electron temperature thus remains at $T_{\rm e} > 10^9$ K in the cocoon.
X-ray emissions from the cocoon are weak because the electron number density is low.
Meanwhile, soft X-rays are emitted from the shocked intracluster medium surrounding the cocoon.
Mixing of the jet plasma and the shocked intracluster medium through the Kelvin--Helmholtz instability at the interface enhances X-ray emissions around the contact discontinuity between the cocoon and shocked intracluster medium.
\end{abstract}




\section{Introduction}
Powerful jets ejected from active galactic nuclei (AGNs) interact with the intracluster medium (ICM).
Hot spots observed in AGN jets indicate that the bulk kinetic energy of the jets is converted to the thermal energy of ions and electrons as well as the energy of nonthermal particles.
The hot spots are most likely jet terminal shocks, through which the jet plasma warms, expands, and flows back toward the galactic nucleus.
This backflow is referred to as a cocoon \citep[e.g.,][]{1974MNRAS.169..395B,1974MNRAS.166..513S,1989ApJ...345L..21B}.
The hot spots and cocoon are bright in radio emissions because nonthermal electrons generated at the terminal shock emit synchrotron radiation.
However, it is difficult to detect the thermal emission from the cocoon because the electron number density is low.
Instead, the cocoon is observed as an X-ray cavity \citep{2000MNRAS.318L..65F}.
Numerical simulations have revealed that structures similar to X-ray cavities form when the kinetic--energy-dominated light jet interacts with the denser ambient medium  \citep[e.g.,][]{1992PASJ...44..245T,1997ApJ...479..151M,1999ApJ...523L.125A,2010MNRAS.402....7M,2014MNRAS.445.1462P,2005A&A...431...45K,2006MNRAS.373L..65H,2010ApJ...725.1440M,2011ApJ...728..121G}.
These simulations indicated that the X-ray cavity and backflow affect the ICM as a reservoir of thermal energy.

Astrophysical plasma such as that in AGN jets, galaxy clusters, and supernova remnants are almost collisionless; i.e., the collisional mean free path is longer than their size.
In the absence of collisions that would enforce thermal equilibrium, electrons and ions do not always have the same temperature.
Moreover, shocks primarily heat ions because the kinetic energies of electrons and ions are proportional to the masses of these particles.
Observational evidence of two-temperature plasma has been obtained for various astrophysical phenomena, such as the bow shocks of the Earth and Saturn \citep{1988JGR....9312923S,2011JGRA..11610107M}, the forward shock of supernova remnants \citep{2013SSRv..178..633G}, and merging galaxy clusters \citep{2012MNRAS.423..236R}.
Observations of the bow shocks of the Earth and Saturn indicate that the ratio of the electron temperature to ion temperature $T_{\rm e}/T_{\rm i}$ is unity when ${\mathcal M} < 3$, where ${\mathcal M}$ is the Mach number, and
$T_{\rm e}/T_{\rm i} \propto {\mathcal M}^{-2}$\citep{2013SSRv..178..633G} otherwise.
The post-shock region of galaxy clusters requires a two-temperature treatment.
\cite{2012MNRAS.423..236R} showed that, for the galaxy cluster Abell 2146, the post-shock electron temperature is lower than the ion temperature, as predicted from Rankine--Hugoniot jump conditions.

Jets of low-luminosity AGNs are launched from radiatively inefficient accretion flows.
The collision time scale is much longer than the accretion time scale, and the electron temperature is thus lower than the ion temperature\citep[e.g.,][]{1976ApJ...204..187S,1995ApJ...452..710N,1996PASJ...48..761N}.
\cite{1997ApJ...489..791M} calculated the global structure of an advection-dominated accretion flow comprising two-temperature plasma and obtained a model that explains the spectrum of Sagittarius A*, where the temperature profile of the electron is the vital factor.
\cite{2015MNRAS.454.1848R} reported the results of simulations of a two-temperature accretion disk carried out by combining the electron energy equation with ideal general-relativistic magnetohydrodynamics (MHD) equations.
The two-temperature MHD approach has been applied in recent simulations of the jet formation in M87 \citep{2018ApJ...864..126R,2019MNRAS.486.2873C}.

A key factor of the two-temperature MHD is the fraction of the electron heating to the dissipated energy.
The ratio of electron heating to ion heating depends on the microscopic properties of the collisionless plasma on a scale much smaller than the cell size in MHD simulation.
Previous studies assumed two different physical mechanisms in estimating the fraction of electron heating \citep{2018MNRAS.478.5209C,2019MNRAS.486.2873C}.
The first mechanism is MHD turbulent heating in weakly collisional plasma while the second is heating by magnetic reconnection.
Results of gyrokinetic simulations for turbulent heating \citep{2010MNRAS.409L.104H} indicate that the heating rate strongly depends on the ratio of the ion pressure to the magnetic pressure $\beta_{\rm i}$.
Almost all energy dissipated by turbulence goes to electrons in the low-$\beta_{\rm i}$ region.
When electrons are heated by turbulent heating, the electron temperature becomes one order of magnitude higher than the ion temperature in the funnel region of the jets where $\beta_{\rm i}$ is low.

When the electron is heated by fast magnetic reconnection, the fraction of the electron heating does not exceed 0.5.
This means that the electron always obtains less dissipated energy than the ion.
This tendency has been observed in particle-in-cell simulations of fast magnetic reconnection \citep{2017ApJ...850...29R}.
The electron temperature is therefore equal to or less than the ion temperature in the funnel region of the jets.
These simulation results indicate the existence of a two-temperature plasma in jets.
However, it is not obvious whether electrons and ions have different temperatures during the propagation of large-scale AGN jets.

In our previous work \cite{2019Galax...7...14O}, we investigated the propagation of sub-relativistic jets in X-ray binaries by conducting two-temperature MHD simulations.
We showed  that the ion temperature downstream of the jet terminal shock (hot spot) becomes 10 times the electron temperature because ions are heated by energy dissipation at shocks.
Meanwhile, electrons are not heated at the shock front because the instantaneous electron heating at shock fronts was ignored.
In the cocoon, electrons are heated by Coulomb collisions with ions. Around the interface between the cocoon and the ambient medium, the electron temperature decreases owing to gas mixing between the hot cocoon plasma and the low-temperature ambient plasma via Kelvin--Helmholtz (KH) instability.

In the present paper, we examine the effect of instantaneous heating on the electron temperature distribution and the dependence of electron and ion temperatures on initial conditions.
We report the results of two-temperature MHD simulations of a sub-relativistic AGN jet that propagates and interacts with the ICM, including Coulomb coupling and instantaneous electron heating at shocks for the first time.
The remainder of the paper is organized as follows.
Section \ref{sec:method} describes basic equations for the two-temperature and single-fluid MHD simulation and the numerical setup for two-dimensional simulations.
In particular, section \ref{sec:method-dis} explains the method of calculating dissipation heating.
Sections \ref{sec:results} and \ref{sec:discussion} respectively present and discuss our results.
Section \ref{sec:summary} presents conclusions.

\section{Numerical Method}
\label{sec:method}
\subsection{Basic Equations}
The basic equations are two-temperature single-fluid MHD equations.
We assume fully ionized hydrogen plasma and charge neutrality $n=n_{\rm i}=n_{\rm e}$, where $n_{\rm i}$ and $n_{\rm e}$ are respectively the ion number density and electron number density.
The single-fluid two-temperature MHD equations are
\begin{equation}
   \label{eq:mhd1}
   \frac{\partial n}{\partial t} + \nabla \cdot (n \bm{v}) = 0,
\end{equation}
\begin{equation}
   \label{eq:mhd2}
   m_{\rm i}n \left[\frac{\partial\bm{v}}{\partial t}+(\bm{v}\cdot \nabla)\bm{v} \right] =- \nabla p_{{\rm gas}} - \nabla \left( \frac{B^2}{8\pi}\right)+\frac{1}{4\pi} \left( \bm{B} \cdot \nabla \right)\bm{B},
\end{equation}
\begin{equation}
   \label{eq:mhd3}
   \frac{\partial \bm{B}}{\partial t} = \nabla \times (\bm{v} \times \bm{B}),
\end{equation}
\begin{equation}
  \label{eq:mhd4}
   \frac{\partial \epsilon_{\rm i}}{\partial t}+\nabla \cdot [(\epsilon_{\rm i}+p_{\rm i})\bm{v}] -(\bm{v}\cdot\nabla)p_{\rm i}
   = - q^{\rm ie} + (1-f_{\rm e}) q^{\rm heat},
\end{equation}
\begin{equation}
  \label{eq:mhd5}
      \frac{\partial \epsilon_{\rm e}}{\partial t}+\nabla \cdot [(\epsilon_{\rm e}+p_{\rm e})\bm{v}] -(\bm{v}\cdot\nabla)p_{\rm e}
   = + q^{\rm ie} + f_{\rm e}q^{\rm heat},
\end{equation}
where $\bm{v}$ is the velocity, $\bm{B}$ is the magnetic field, $m_{\rm i}$ is the ion mass, $p_{{\rm gas}}=p_{\rm i}+p_{\rm e}$ is the gas pressure, and $p_{\rm i}$ and $p_{\rm e}$ are respectively the ion and electron gas pressures.
We assume an ideal gas, and the internal energies of ions and electrons are thus
\begin{equation}
  \epsilon_{\rm i}= \frac{p_{\rm i} }{\gamma_{\rm i}-1},~~\epsilon_{\rm e}= \frac{p_{\rm e} }{\gamma_{\rm e}-1},
\end{equation}
where $\gamma_{\rm i}$ and $\gamma_{\rm e}$ are respectively the specific-heat ratios for ions and electrons.
We consider two types of heating source in the energy equation of the electron.
One is $q^{\rm ie}$, which is the rate of energy transfer from ions to electrons through Coulomb coupling.
$q^{\rm heat}$ is a dissipative heating rate.
Here $f_{\rm e}$ is the fraction of electron heating.
We ignore radiative cooling in this paper.

Ion and electron temperatures are thus given by the ideal equation of state,
\begin{equation}
  p_{\rm i} = nk_{\rm B}T_{\rm i},~~p_{\rm e} = nk_{\rm B}T_{\rm e},
\end{equation}
where $k_{\rm B}$ is the Boltzmann constant.
The effective temperature of the electron and ion mixed gas is
\begin{equation}
  p_{\rm gas} = (\gamma_{\rm gas}-1)\epsilon_{\rm gas} = 2nk_{\rm B}T_{\rm gas},
\end{equation}
where $\epsilon_{\rm gas}$ is the gas internal energy.
For simplicity, we assume a constant value for $\gamma_{\rm e}=\gamma_{\rm i}=\gamma_{\rm gas} = 5/3$.

The rate of energy transfer, $q^{{\rm ie}}$, from ions to electrons per unit volume through Coulomb collisions is \citep{1983MNRAS.204.1269S,1991ApJ...369..410D}
\begin{eqnarray}
 \label{eq:qie}
 q^{{\rm ie}} &=& \frac{3}{2} \frac{m_{\rm e}}{m_{\rm i}} n^2 \sigma_{\rm T} c
 \frac{\ln{\Lambda} (k_{\rm B} T_{\rm i} - k_{\rm B} T_{\rm e})}{K_2\left(1/\theta_{\rm e} \right)K_2\left( 1/\theta_{\rm i} \right)  } \nonumber \\
  &\times&
  \left[
  \frac{2(\theta_{\rm e} + \theta_{\rm i})^2 +1}{\theta_{\rm i} + \theta_{\rm e}}K_1\left(\frac{1}{\theta_{\rm m}}\right)
  + 2K_0\left(\frac{1}{\theta_{\rm m}} \right)
  \right]  ~~~~(\theta_{\rm i} > 0.2) ,\nonumber \\
 q^{{\rm ie}} &=& \frac{3}{2} \frac{m_{\rm e}}{m_{\rm i}} n^2 \sigma_{\rm T} c \ln{\Lambda} (k_{\rm B} T_{\rm i} - k_{\rm B} T_{\rm e}) \nonumber \\
 &\times&\frac{\sqrt{\pi/2}+\sqrt{\theta_{\rm i}+\theta_{\rm e}}}{(\theta_{\rm i}+\theta_{\rm e})^{3/2}}  ~~~~(\theta_{\rm i} < 0.2),
\end{eqnarray}
with $\theta_{\rm m}=\theta_{\rm i} \theta_{\rm e}/(\theta_{\rm i}+\theta_{\rm e})$.
The parameters $\sigma_{\rm T}$, $c$, and $\ln \Lambda$ are respectively the Thomson scattering cross section, the speed of light, and the Coulomb logarithm, which is 20.
Functions $K_{0}, K_{1}$, and $K_{2}$ are respectively modified Bessel functions of the second kind of order 0, 1, and 2.
The quantities
\begin{equation}
  \theta_{\rm i}=\frac{k_{\rm B}T_{\rm i}}{m_{\rm i} c^2}~~~{\rm and}~~\theta_{\rm e}=\frac{k_{\rm B}T_{\rm e}}{m_{\rm e}c^2}
\end{equation}
are respectively the dimensionless ion and electron temperatures.
The units of velocity and length are respectively $v_0 = 9.09\times10^7$ cm/s and $r_{\rm jet} =1$ kpc, corresponding to the jet radius.
The normalized density is $\rho_0 = 0.835\times10^{-25} \alpha$ g/cc.
Here, $\alpha$ is a density parameter used to study the effects of Coulomb coupling, and we set $\alpha = 1,10^{-1},10^{-2}$.
Other numerical parameters are listed in Table \ref{tab:normalized}.
\begin{table}
  \begin{center}
  \caption{Units adopted in this paper}
  \label{tab:normalized}
    \begin{tabular}{c c c}
      \hline
                  & Numerical Unit & Physical Unit \\ \hline
      length      & $r_{\rm jet}$  & 1kpc\\
      velocity    & $v_{\rm 0}$    & $9.09\times10^{7}$cm/s\\
      time        & $t_{\rm 0}$    & 1Myr\\
      density     & $\rho_{\rm 0}$ & $ 0.835\times10^{-25} \alpha$ g/cc \\
      temperature & $T_{\rm 0}$    & $1.0\times10^{8}$K\\
      \hline
  \end{tabular}
\end{center}
\end{table}
\subsection{Numerical Scheme}
We modified the MHD code CANS+ \citep{2019PASJ...71...83M} to include the energy equation for electrons.
CANS+ solves the Newtonian--MHD equations in conservation form as follows.
(1) Reconstruction adopts a fifth-order monotonicity-preserving interpolation scheme \citep{1997JCoPh.136...83S}.
(2) The time integral is performed with third-order total-variation-diminishing preserving Runge--Kutta methods.
(3) The numerical flux across cell interfaces is computed using the HLLD Riemann solver \citep{2005JCoPh.208..315M}.
(4) The hyperbolic divergence cleaning method is adopted for a magnetic field \citep{2002JCoPh.175..645D}.

Our numerical code solves the equations in conservation form.
We therefore arrange the two-temperature MHD equations (eqs. \ref{eq:mhd1}--\ref{eq:mhd5}) in conservation form:
\begin{equation}
  \label{eq:conserved}
  \frac{\partial{\bm{U}}}{\partial{t}} +\nabla \cdot \bm{F} = \bm{S},
\end{equation}
\begin{equation}
  \bm{U} = \left(
             \begin{array}{c}
                n        \\
                m_{\rm i}n \bm{v} \\
                \bm{B}      \\
                e           \\
                n \kappa_{\rm e}    \\
             \end{array}
           \right) ,
\end{equation}
\begin{equation}
  \bm{F} = \left(
             \begin{array}{c}
              n \bm{v}  \\
              m_{\rm i}n \bm{v} \bm{v} + p_{\rm T} \bm{I} - \frac{1}{4\pi}\bm{B}\bm{B}  \\
               \bm{v}\bm{B}-\bm{B}\bm{v} \\
               (e+p_{\rm T})\bm{v}-\frac{1}{4\pi}\bm{B}(\bm{v}\cdot \bm{B}) \\
               n \kappa_{\rm e} \bm{v}
             \end{array}
           \right)  ,
\end{equation}
\begin{equation}
  \bm{S} = \left(
             \begin{array}{c}
               0 \\
               0  \\
               0 \\
               0 \\
               (\gamma_{\rm e}-1)n^{1-\gamma_{\rm e}} (q^{\rm ie}+f_{\rm e}q^{\rm heat})
             \end{array}
           \right)  ,
\end{equation}
where $\bm{U}$ and $\bm{F}$ respectively denote the conserved quantities and flux vectors, $\bm{I}$ is a unit matrix, $\bm{S}$ is the source term, and the total energy, $e$, is given by
\begin{equation}
  e =  \epsilon_{\rm i} + \epsilon_{\rm e} + \frac{1}{2}\rho \bm{v}^2 + \frac{1}{8\pi}\bm{B}^2.
\end{equation}
$p_{\rm T} = p_{\rm i} + p_{\rm e} + p_{\rm mag}$ is total pressure.
We have here used the electron pseudo-entropy $\kappa_{\rm e} \equiv p_{\rm e}n^{-\gamma_{\rm e}}$.
The electron gas entropy per particle is given by
\begin{equation}
   s_{\rm e} = k_{\rm B}(\gamma_{\rm e}-1)^{-1} \log{(p_{\rm e}n^{-\gamma_{\rm e}})} = k_{\rm B}(\gamma_{\rm e}-1)^{-1} \log{\kappa_{\rm e}}.
\end{equation}
The source term is updated implicitly by adopting the Newton--Raphson iteration.

\subsection{Calculation of Dissipation Heating}
\label{sec:method-dis}
To calculate the dissipation heating rate
\footnote{In our simulations, we solve ideal MHD equations, and numerical viscosity is the origin of dissipation.}
 $q^{\rm heat}$, we adopt an approach similar to that followed by \cite{2015MNRAS.454.1848R} and \cite{2017MNRAS.466..705S}.
We adopt the following to evaluate the dissipated energy at each time step.

1. We solve the conserved equations (\ref{eq:conserved}), and obtain the gas specific internal energy $\epsilon_{\rm gas}$ at time step $n+1$:
\begin{equation}
  \epsilon^{n+1}_{\rm gas} = \frac{p^{n+1}_{\rm gas}}{\gamma_{\rm gas}-1}.
\end{equation}

2. To compute the purely adiabatic evolution, we use the gas entropy conservation equation:
\begin{equation}
      \frac{\partial}{\partial t}\left( n\kappa_{\rm gas} \right) + \nabla \cdot \left( n\kappa_{\rm gas} \bm{v} \right) = 0,
\end{equation}
where $\kappa_{\rm gas}$ is the gas pseudo-entropy.
To solve the above equation, we solve the finite difference equation between the $n$-th and $(n+1)$-th time step for each cell and adopt the fifth-order monotonicity-preserving method.
The gas specific thermal energy that evolves under the adiabatic process is then calculated as
\begin{equation}
  \epsilon^{n+1}_{\rm gas,ad} = \frac{ \left( \kappa_{\rm gas} n^{\gamma_{\rm e}} \right)^{n+1} }{ \gamma_{\rm e}-1 }.
\end{equation}

3. The dissipation heating rate is therefore estimated as
\begin{equation}
  q^{\rm heat} = \frac{\epsilon^{n+1}_{\rm gas} -\epsilon^{n+1}_{\rm gas,ad}}{\Delta t}.
  \label{eq:dissipated heat}
\end{equation}
\subsection{Numerical Setup}
We present the results of 10 models used to investigate the effect of electron heating on the electron temperature distribution and the dependence of electron and ion temperatures on the initial conditions.
We perform axisymmetric simulations of the large-scale jet evolution when the jet is injected into a medium having constant density.

Parameters common to all models are summarized in Table \ref{tab:common}.
The computational domain is $0 < r/r_{\rm jet} < 40$, $0 < z/r_{\rm jet} <80$ and the number of numerical cells is $(N_{\rm r},N_{\rm z}) = (1024,2048)$.
The initial radius of the jet is 1 kpc, resolved by 24 numerical cells.
The initial Mach number of jets ${\mathcal M}_{\rm jet}$ is 14, and the ratio of the thermal pressure of the jet to the ICM is 10.
The speed of the injected jet is 0.2c.
The components of the injected magnetic field are
\begin{eqnarray}
 B_{\rm \phi} &=& \begin{cases}
                B_{\rm in} \sin^4{(2\pi r/r_{\rm jet})} & (r<r_{\rm jet}) \\
                0 & ({\rm otherwise}) \nonumber
                \end{cases}\\
 B_{\rm r} &=& B_{\rm z} = 0,
\end{eqnarray}
where $B_{\rm in}$ is given by the plasma beta as $B_{\rm in} = (8\pi p_{\rm gas}/\beta)^{1/2}$, with $\beta=p_{\rm gas}/p_{\rm mag}$.
The temperature of the injected gas is $T_{\rm gas} = 0.5(T_{\rm e}+T_{\rm i}) = 1.0\times10^9$ K.
The initial ICM is unmagnetized, and the density ratio of the jet beam to the ICM  ($\eta \equiv \rho_{\rm jet}/\rho_{\rm ICM}$) is $10^{-2}$.

Table \ref{tab:model} presents our numerical models.
We adopt three values of the fraction of electron heating, $f_{\rm e}= 0.0, 0.05, 0.2$, in studying the effect of electron heating.
The ratio of injected electron and ion temperatures is set at $m \equiv T_{\rm e,inj}/T_{\rm gas,inj} = 1.9, 1.0, 0.1$ to examine the dependence on the initial condition.
Moreover, we switch on and off Coulomb coupling.
\begin{table}
  \begin{center}
  \caption{Common simulation setup parameters}
  \label{tab:common}
    \begin{tabular}{c c c}
      \hline
      Jet speed               & $v_{\rm jet}$   & 0.2c \\
      Jet gas temperature     & $T_{\rm g,jet}$ & $1.0\times10^9$ K\\
      Jet plasma $\beta$      & $\beta$ & 10 \\
      Jet Mach Number         & ${\mathcal M}_{\rm jet}$ &14 \\
      Ambient gas density     & $\rho_{\rm ICM}$ & $0.835\times10^{-24}\alpha$\\
      Ambient gas temperature & $T_{\rm ICM}$ & $1.0\times10^6$ K\\
      Density ratio           & $\rho_{\rm jet}/\rho_{\rm ICM})$ & $10^{-2}$\\
      Gas pressure ratio      & $p_{\rm jet}/p_{\rm ICM} $ & $10$ \\
      \hline
  \end{tabular}
\end{center}
\end{table}
\begin{table}
  \begin{center}
  \caption{Simulation models.
  The columns give the model name, ratio of the injection electron temperature to the gas temperature, the fraction of the electron heating, Coulomb coupling, and normalized density parameter.}
  \label{tab:model}
    \begin{tabular}{c c c c c}
      \hline
      Model     &$m \equiv T_{\rm e,inj}/T_{\rm gas,inj}$ & $f_{\rm e}$ & Coulomb coupling & $\alpha$   \\ \hline
      f00m1     &   1                             & 0.0         & -                & 1          \\
      f00m1C    &   1                             & 0.0         & ON               & 1          \\
      f005m1    &   1                             & 0.05        & -                & 1          \\
      f005m1C   &   1                             & 0.05        & ON               & 1          \\
      f02m1     &   1                             & 0.2         & -                & 1          \\
      f02m1C    &   1                             & 0.2         & ON               & 1          \\
      f02m0.1C    &   0.1                         & 0.2         & ON               & 1          \\
      f02m1.9C    &   1.9                         & 0.2         & ON               & 1          \\
      f005m1C$\alpha$-1 &   1                     & 0.05        & ON               & $10^{-1}$  \\
      f005m1C$\alpha$-2 &   1                     & 0.05        & ON               & $10^{-2}$  \\ \hline
  \end{tabular}
\end{center}
\end{table}
\section{Numerical Results}
\label{sec:results}
In Appendix \ref{sec:1d}, we show the result of the one-dimensional Riemann problem for our jet model to examine the electron temperature dependence on the fraction of the electron heating $f_{\rm e}$ at shock fronts.
We find that the post-shock temperature ratio of the electron to ion is simply described by eq. (\ref{eq:post-shock temp}).
In this section, we investigate the multidimensional effects and the dependence on the temperatures of the injected electrons and ions.
In addition, we present the time evolution to clarify when and where Coulomb coupling is effective.
\subsection{Morphology and Temperature Distribution}
\begin{figure*}
	\includegraphics[width=0.95\textwidth]{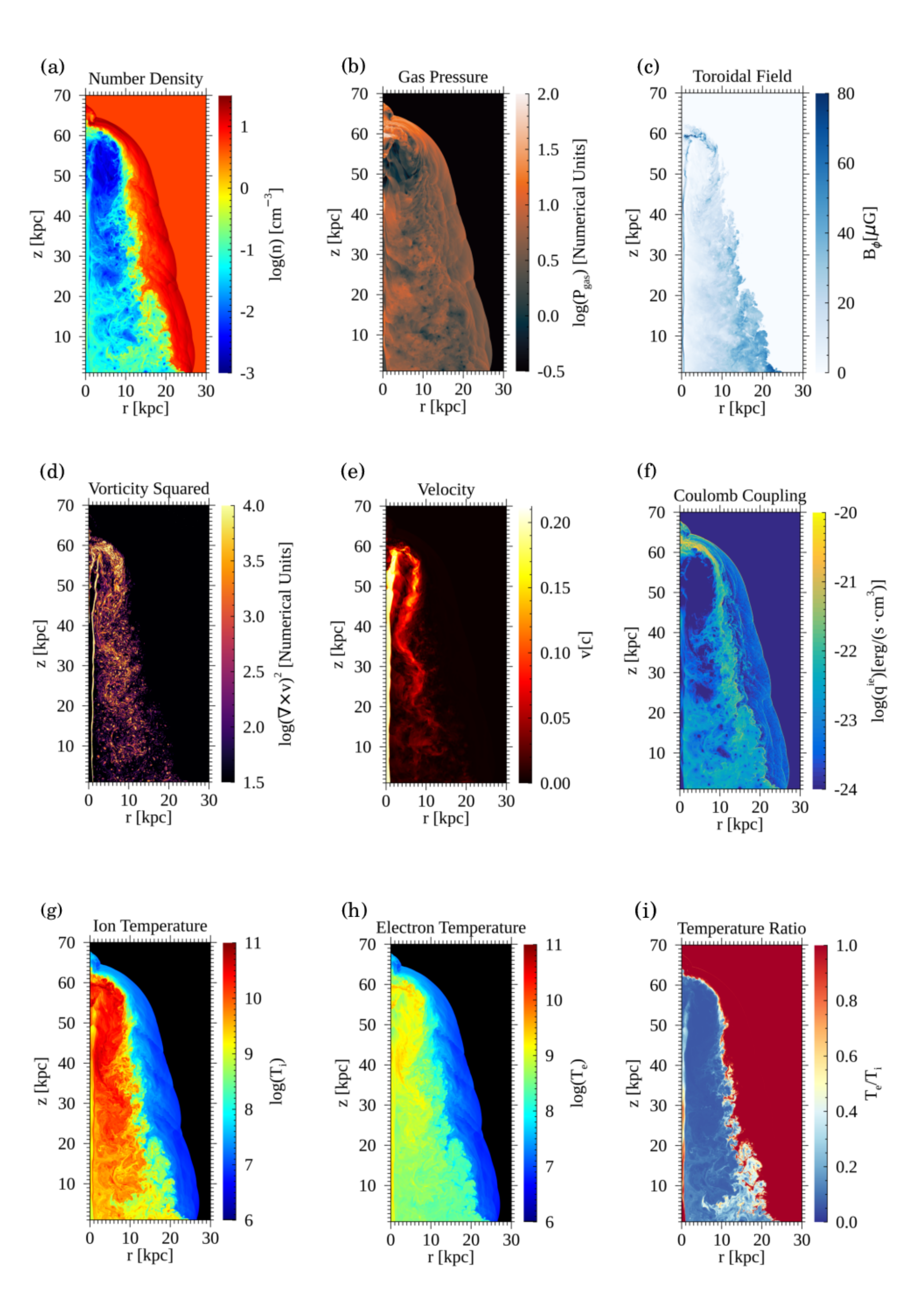}
    \caption{Snapshots of (a) the number density, (b) the gas pressure, (c) the toroidal magnetic field, (d) the vorticity squared $(\nabla \times \bm{v})^2$, (e) the flow absolute velocity, (f) the energy transfer rate from ions to electrons through Coulomb coupling, (g) the ion temperature, (h) the electron temperature, and (i) the temperature ratio of electrons to ions for model f005m1C ($f_{\rm e}=0.05$) at $t=20.0$ Myr.
    }
    \label{fig:Fiducial}
\end{figure*}

This subsection presents the results of a fiducial model of jet propagation.
The fiducial model is the model f005m1C.
Figure  \ref{fig:Fiducial} shows snapshots of (a) the number density, (b) the gas pressure, (c) the toroidal magnetic field component, (d) the vorticity squared, and (e) the absolute velocity at $t=20.0$ Myr.
When the jet propagates into the ICM, several types of shock front form; e.g., internal shocks (i.e., recollimation shocks), the terminal shock (i.e., reverse shock), and the bow shock (i.e., forward shock).
In addition, the shocked matter of the terminal shock forms a backflow called a cocoon.
The bow shock compresses the ICM and forms a high-density shell called the shocked-ICM between the contact discontinuity and bow shock.
The kinetic energy dissipates and convert to thermal energy at shocks, and the pressure of the post-shock gas thus becomes 10 to a 100 times that of the pre-shocked gas.

In our jet model, the magnetic energy is much less than the kinetic and thermal energy.
Therefore, the Lorentz force does not have a practical effect on the dynamics and morphology.
When the plasma reaches the terminal shock, the kinetic energy is converted into magnetic energy and gas internal energy.
The backflow generates the vortex motion in the cocoon, and the intensity of the magnetic field increases to about twice that of the injection field.
The toroidal component is not converted into a poloidal component because of axisymmetry and the absence of jet angular motion.
However, \cite{2009MNRAS.400.1785G} showed that toroidal fields are dominant in a cocoon under an axisymmetric condition when the jet has angular momentum.
Note that the poloidal and toroidal fields might easily convert into one another in three-dimensional simulations.

The beam structure depends on the ratio of the jet pressure to the ICM pressure \citep{1982A&A...113..285N}.
Because we choose the pressure ratio to be greater than unity (i.e., the jet beam is under-expanded), the beam has shock diamonds and a sequential structure of compression and expansion.
The beam accelerates to 0.22c, which is $110\%$ of the injection velocity, through the expanding motion.
The beam velocity decelerates through the terminal shock, and the bulk velocity of backflowing plasma is about 0.1c.
KH instabilities develop and form the vortex motion by interaction with the backflowing plasma and shocked-ICM in the cocoon.
The vortex motions drive pressure waves that convert the kinetic energy into the thermal energy in the shocked-ICM through dissipation \citep{2019arXiv190603272B}.
Furthermore, the vortices create high-temperature and low-density spots.

The remaining panels of Fig. \ref{fig:Fiducial} show snapshots of (f) the energy transfer ratio from ions to electrons through Coulomb coupling, (g) the ion temperature, (h) the electron temperature, and (i) the temperature ratio of electrons to ions at $t=20.0$ Myr.
The electron temperature and ion temperature are decoupled due to the heating at the bow shock.
In the shocked ICM, however, ions and electrons are in thermal equilibrium because the relaxation time of Coulomb coupling, which is proportional to the square of the number density, is shorter than the hydrodynamical time scale.
Meanwhile, ion and electron temperatures are separated through internal shocks in the beam.
The post-shock ion and electron temperatures at the terminal shock are about $10^{11}$ and $10^{10}$ K, respectively.
These values are in good agreement with values obtained in the one-dimensional simulation (see Fig. \ref{fig:1djetteti1}).
The collision time scale is longer than the dynamical time scale in the low-density cocoon, and the electron temperature thus remains lower than the ion temperature.
Around the interface between the cocoon and the shocked ICM, the ion and electron temperatures decrease to $10^{8}$ K owing to turbulent mixing.
\subsection{Dependence on the fraction of electron heating }
\begin{figure*}
  \begin{minipage}{0.49\hsize}
    \begin{center}
        \includegraphics[width=0.95\columnwidth]{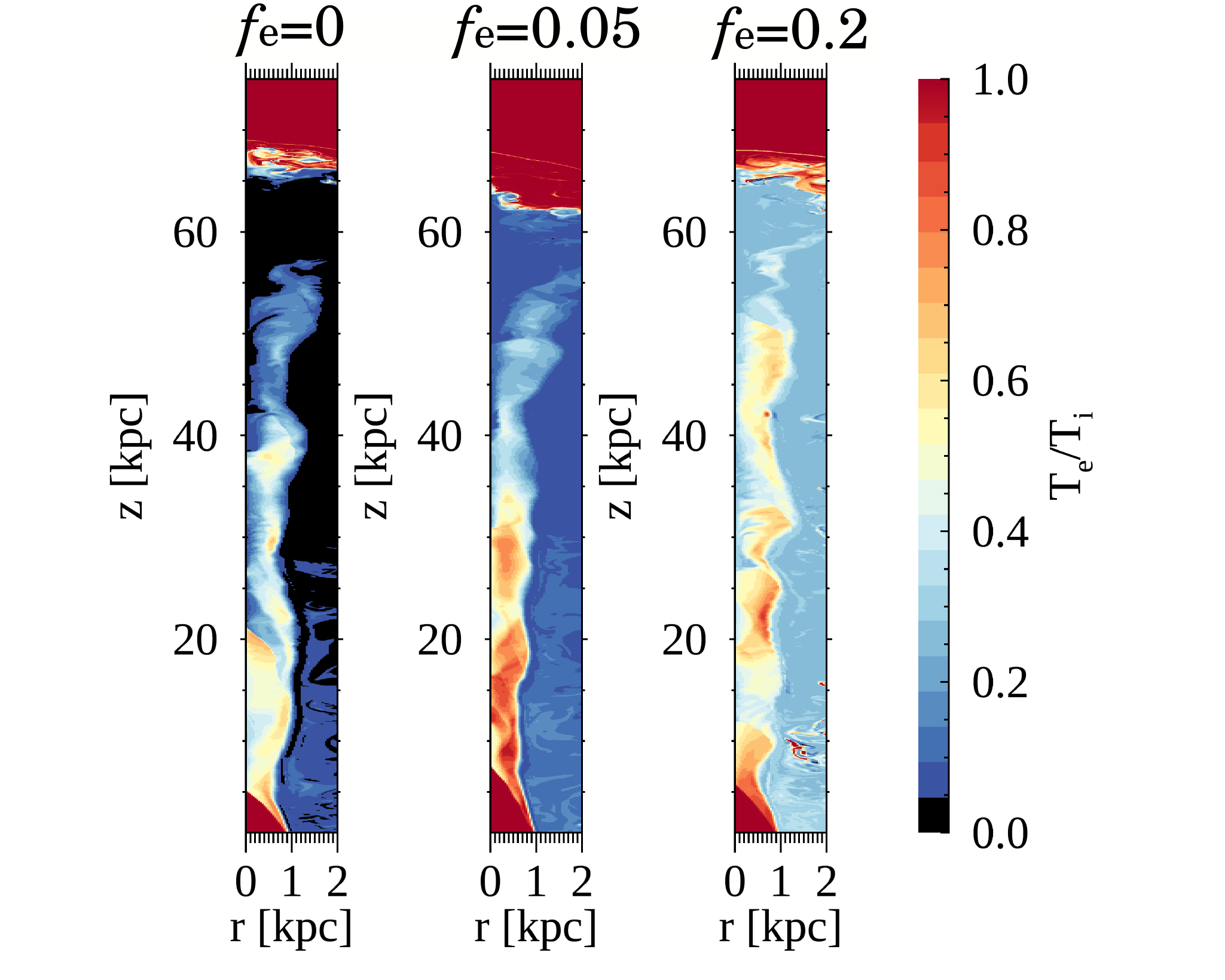}
        \caption{Snapshots of the temperature ratio of electrons to ions in the beam for models f00m1C (left), f005m1C (center), and f02m1C (right).
        }
        \label{fig:fe_depend_map}
    \end{center}
  \end{minipage}
  \begin{minipage}{0.49\hsize}
    \begin{center}
  	\includegraphics[width=0.95\columnwidth]{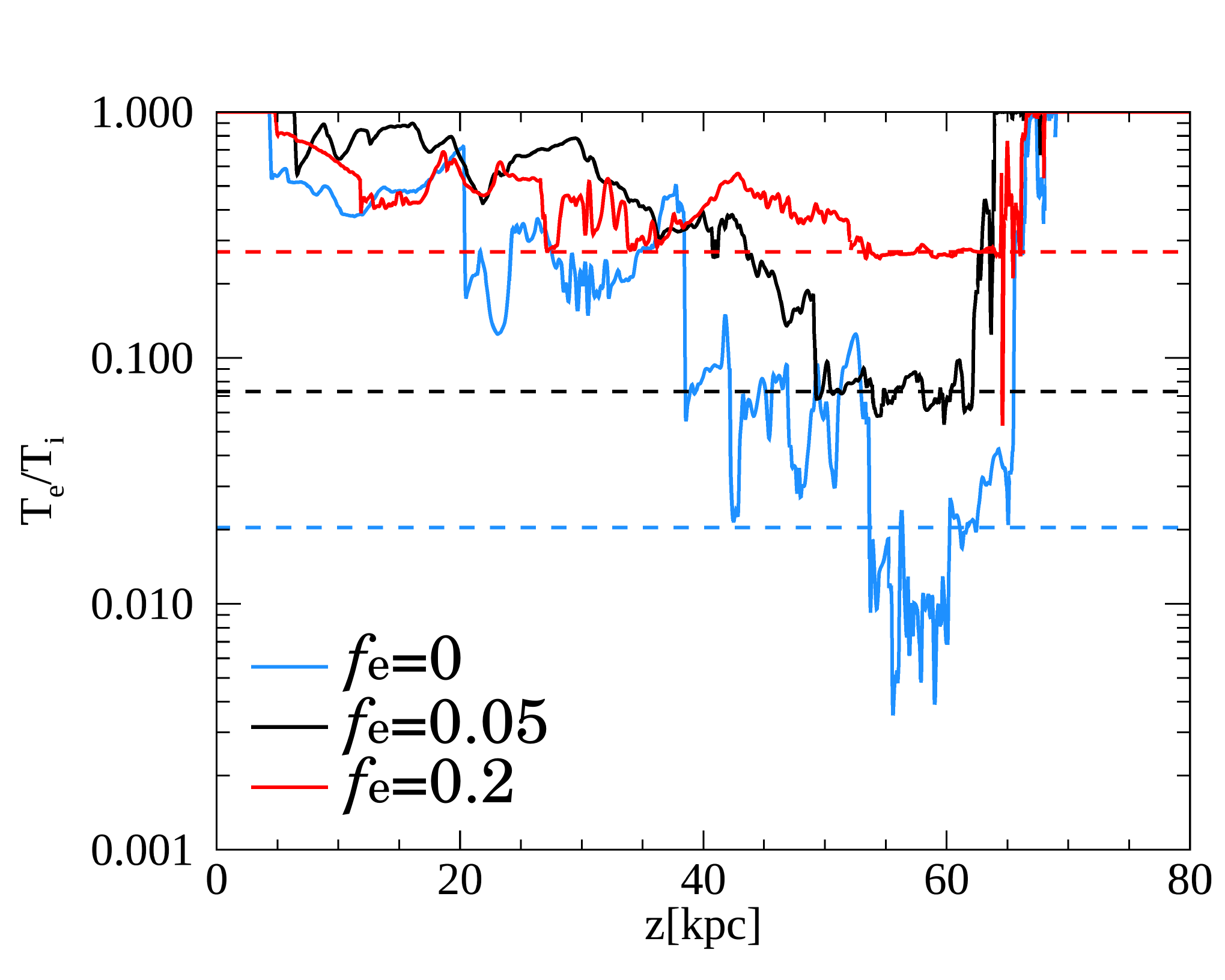}
    \caption{Profiles of the temperature ratio of electrons to ions along the jet beam ($r=0.25$ kpc) for models f00m1C (blue), f005m1C (black), and f02m1C (red).
    The red dashed line and black dashed line respectively show temperature ratios of electrons to ions for $f_{\rm e}=0.2$ and 0.05 predicted using eq. (\ref{eq:post-shock temp}).
    The blue dashed line shows the post-shock temperature ratio when $f_{\rm e}=0$ obtained from eq. (\ref{eq:fe0}).
    }
    \label{fig:fe_depend_1d}
    \end{center}
  \end{minipage}
\end{figure*}

We studied the dependence on the fraction of electron heating $f_{\rm e}$ by carrying out simulations for $f_{\rm e} =0$ (model f00m1C), 0.05 (model f005m1C), and 0.2 (model f02m1C).
Other parameters are the same in these three models.
Figure \ref{fig:fe_depend_map} shows snapshots of the temperature ratio of electrons to ions for models f00m1C (left), f005m1C (center), and f02m1C (right).
Figure \ref{fig:fe_depend_1d} shows the ratio of the electron temperature to the ion temperature along the jet beam ($r=0.25$ kpc) for models f00m1C (blue), f005m1C (black), and f02m1C (red).
The red dashed line and black dashed line respectively show the temperature ratio of the electron to ion for $f_{\rm e}=0.2$ and 0.05 predicted using eq. (\ref{eq:post-shock temp}).
The blue dashed line shows the post--shock temperature ratio when $f_{\rm e}=0$ obtained using eq. (\ref{eq:fe0}).
The ion and electron temperatures separate at the first oblique shock for all models.
The post-shock temperature ratios of electrons and ions at the first oblique shocks are about 0.5 because there remains the effect of electron adiabatic heating.
The temperature ratio approaches the temperature ratio predicted using eq. (\ref{eq:post-shock temp}) when $f_{\rm e} \ne 0$ for each shock (see section \ref{sec:1d}).
The post-shock temperature ratios are close to the predicted values at the terminal shock in the cases that $f_{\rm e} = 0.05$ and 0.2 as gas flows through the beams.
In contrast, the post-shock temperature ratio at the terminal shock is lower than 0.01 for the model f00mC.
This value is smaller than that of the one-dimensional Riemann problem because the electron temperature is reduced by the gas expansion at the hotspot (Fig. \ref{fig:fe_depend_1d} (blue)).
This result indicates that multidimensional effects become important when $f_{\rm e}$ is small.
\subsection{Dependence on Temperatures of Injected Ions and Electrons}
\begin{figure*}
  \begin{minipage}{0.49\hsize}
    \begin{center}
      \includegraphics[width=0.95\columnwidth]{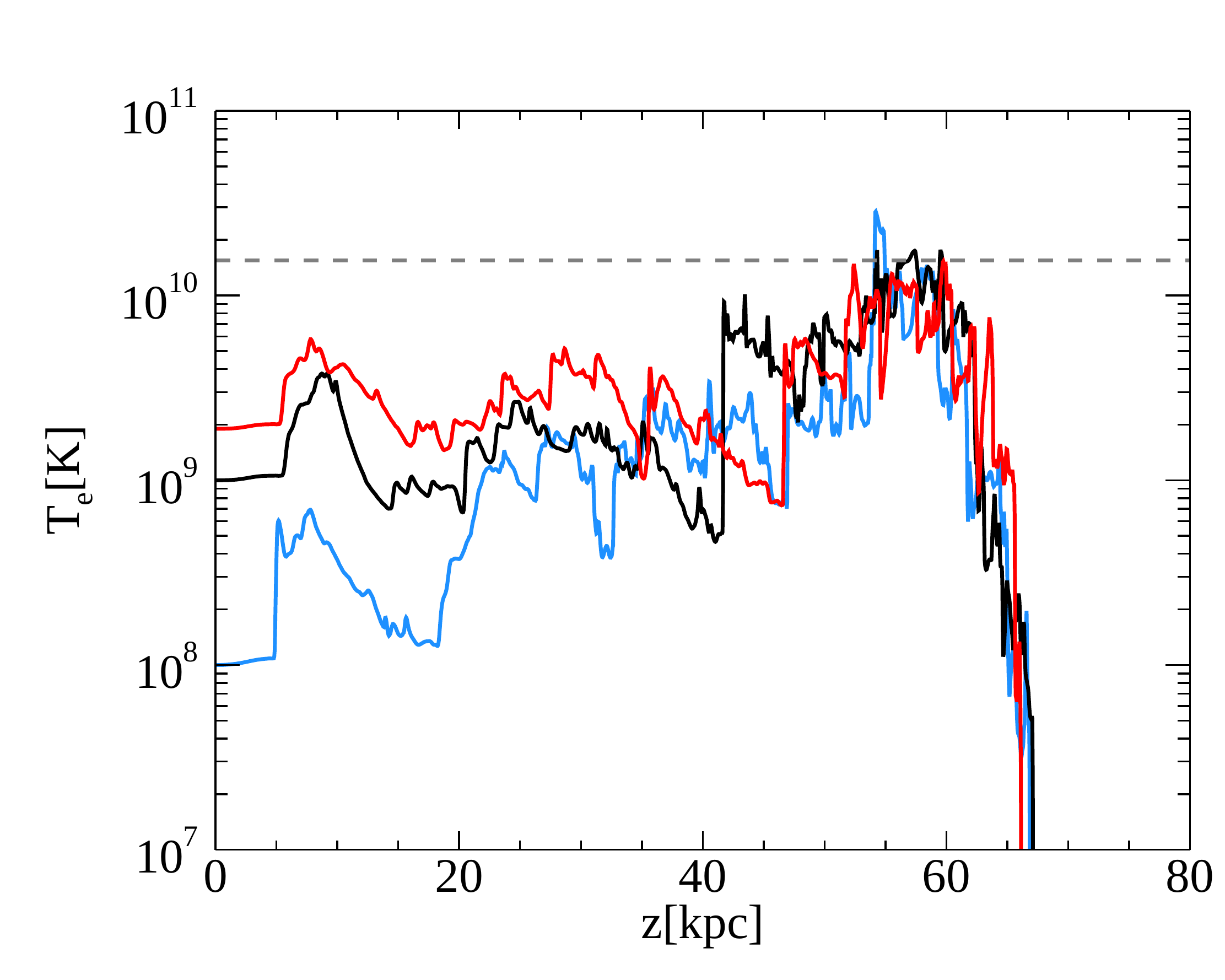}
    \end{center}
  \end{minipage}
  \begin{minipage}{0.49\hsize}
    \begin{center}
      \includegraphics[width=0.95\columnwidth]{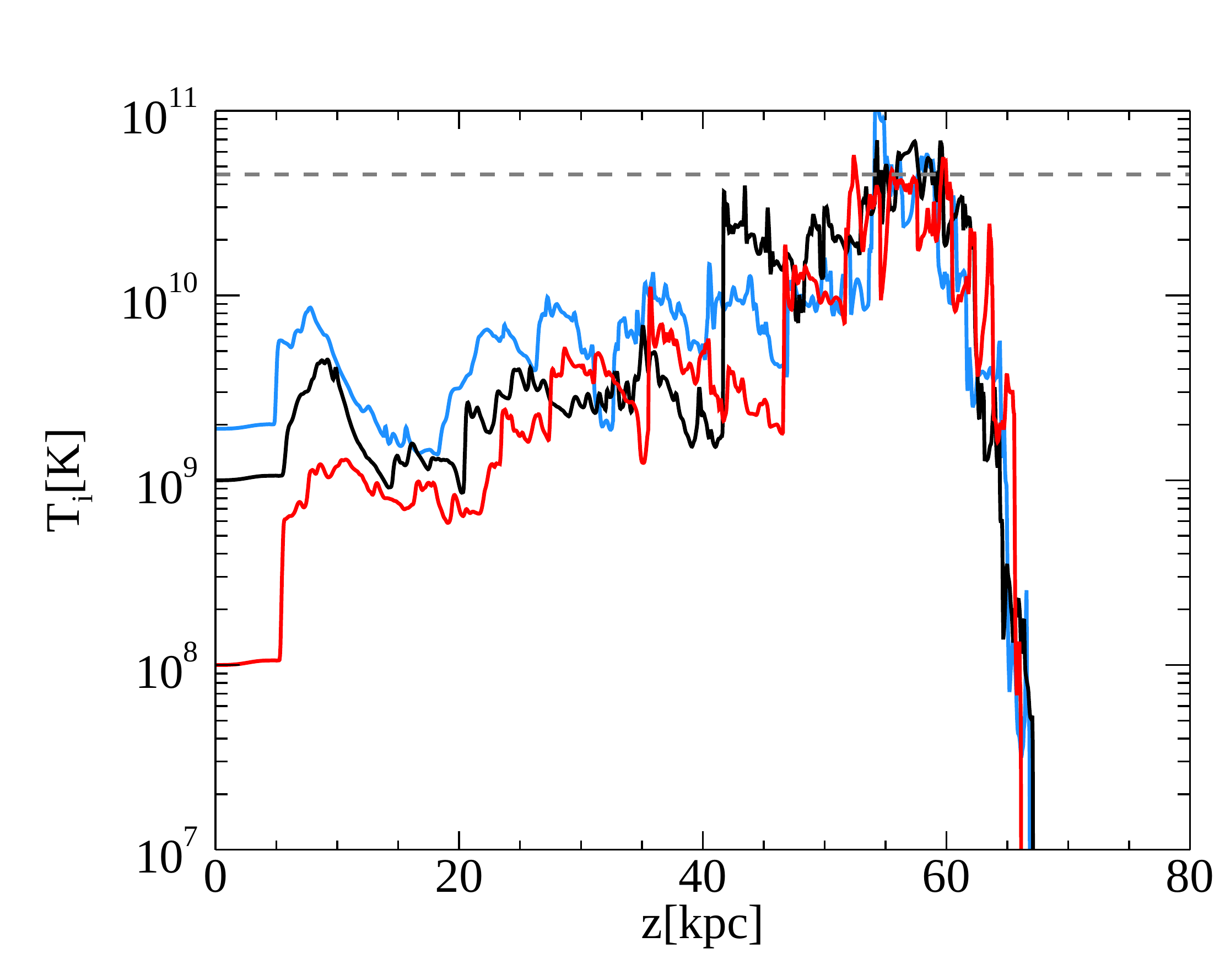}
    \end{center}
  \end{minipage}
    \caption{
    Temperature profiles of the electron (left) and ion (right) along the jet beam at $r=0.25$ kpc for models f02m1C (black), f02m1.9C (red), and f02m0.1C (blue).
    The dashed gray lines show the post-shock electron (left) and ion (right) temperatures predicted using eq. (\ref{eq:post-shock temp}).
    }
    \label{fig:initial}
\end{figure*}

This section compares results for models f02m1C, f02m0.1C, and f02m1.9C to examine the dependence on the temperatures of injected electrons and ions.
These models are the same in that they have $f_{\rm e}=0.2$ with Coulomb coupling but they differ in terms of the temperatures of injected electrons and ions.
Note that the Mach numbers for the plasma are the same in the three models.
Figure \ref{fig:initial} shows the temperature profiles of electron (left) and ion (right) along the jet beam at $r=0.25$ kpc for models f02m1C (black), f02m1.9C (red), and f02m0.1C (blue).
Both electron and ion temperatures increase through the internal shocks.
The electron temperatures of three models are almost the same in the post-shock region of the terminal shock, even though the injection temperatures are different.
This is because the energy dissipated at the terminal shock is much greater than the initial thermal energy.
Therefore, if we know the Mach number of internal jet, we can easily estimate the electron temperature at the hotspot using eq. (\ref{eq:post-shock temp}).
Note that the fraction of electron heating $f_{\rm e}$ could be a function of the temperature ratio.
Meanwhile, when the temperature jump is smaller than the difference between the electron and ion temperatures at a shock (i.e., the Mach number is low or $f_{\rm e}$ is small), the effect of injection must be considered.
The electron temperature at the hotspot and in the cocoon thus depends strongly on the temperature ratio of injected plasma when $f_{\rm e}=0$.

\subsection{Temperature Time Evolution and the Effect of Coulomb Coupling}
\begin{figure}
	\includegraphics[width=0.95\columnwidth]{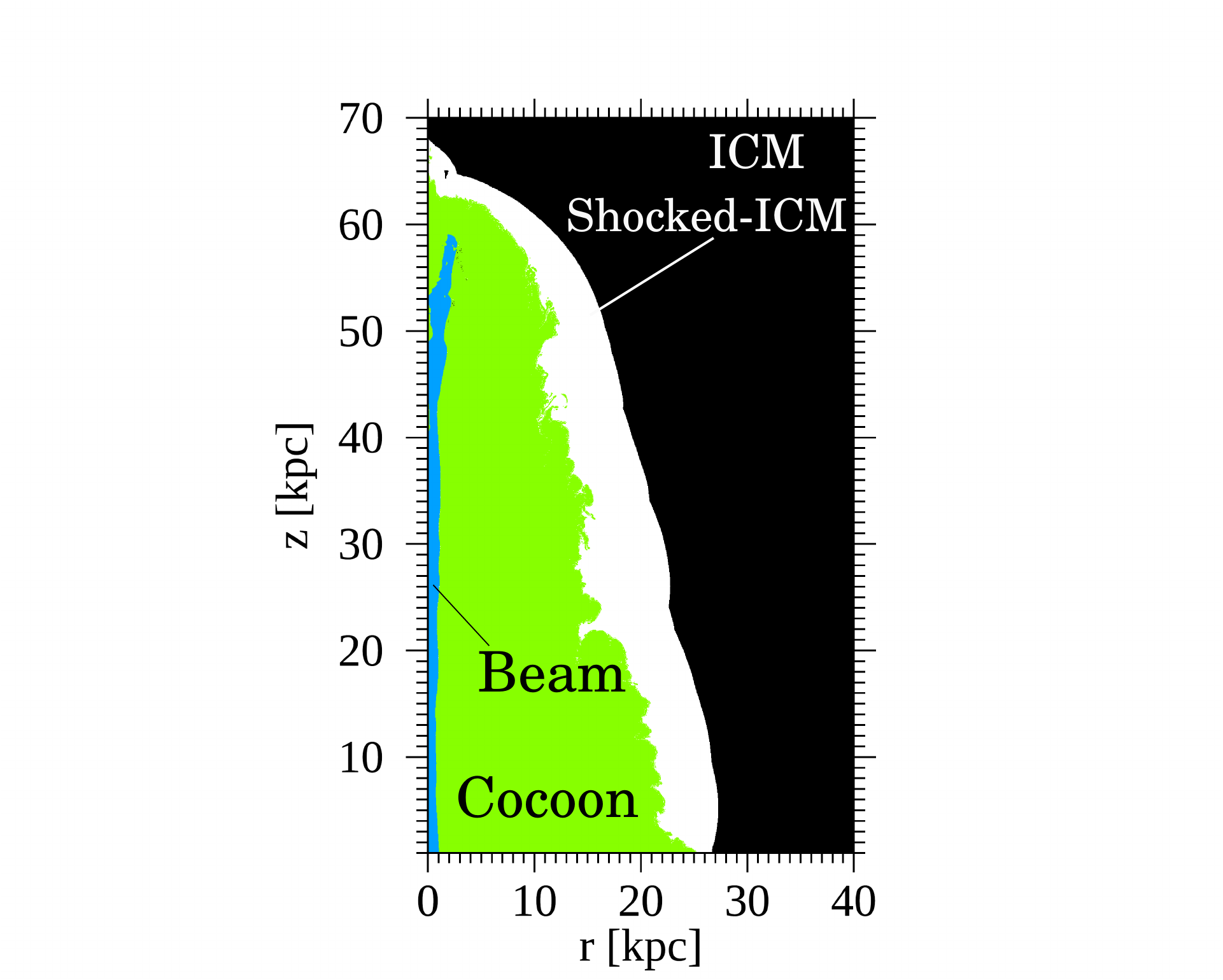}
    \caption{
    Example of divided areas at $t=20.0$ Myr for model f005m1C.
    We define that the beam (blue) is the region where $v_{\rm z} > 0.9v_{\rm z,inj}$ and the cocoon (green) is the region where there is a toroidal magnetic field except for the beam region.
    The shocked ICM (white) has gas pressure higher than its initial value.
    }
    \label{fig:divide}
\end{figure}

\begin{figure}
  \includegraphics[width=0.95\columnwidth]{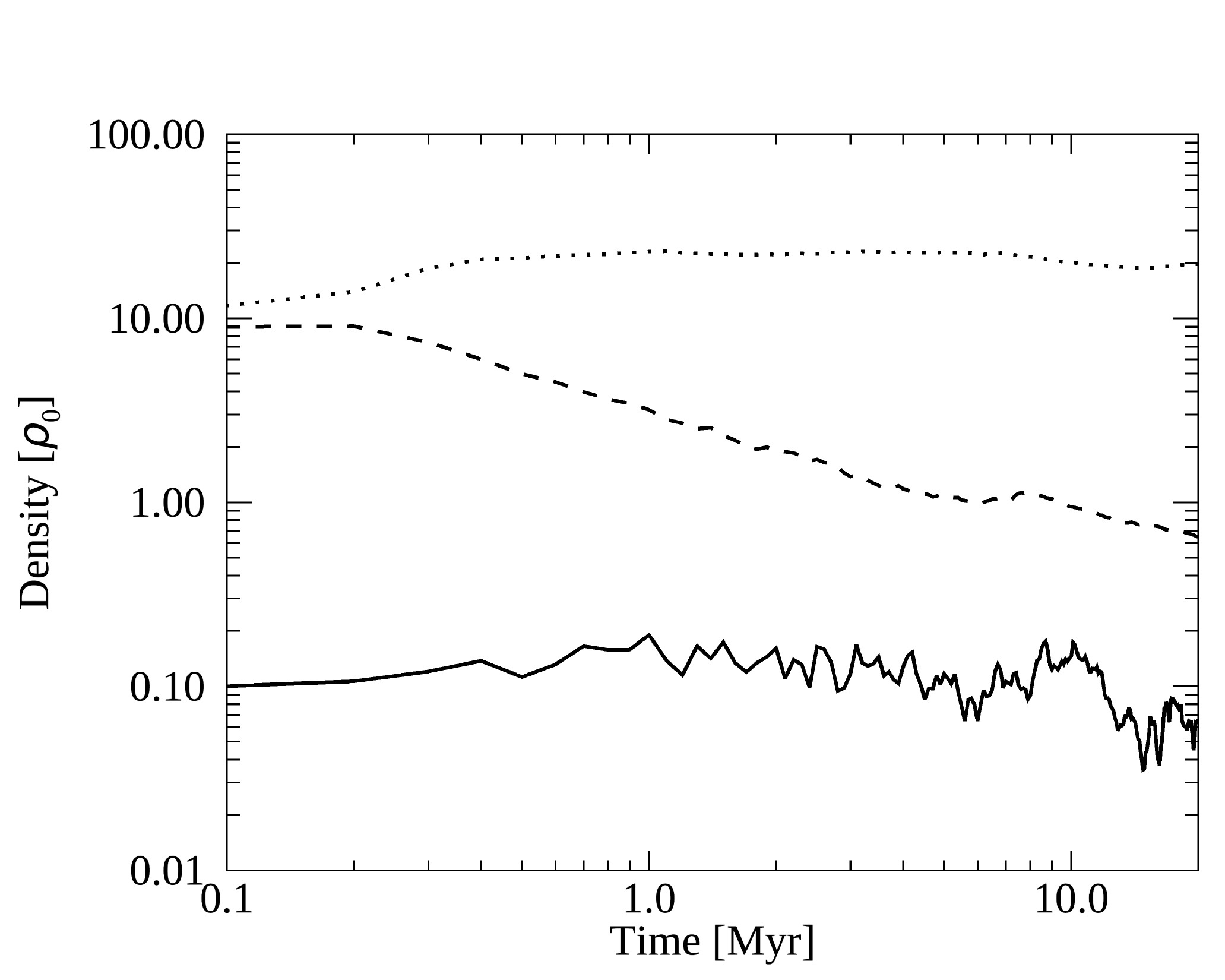}
    \caption{
    Time evolution of the volume-weighted density in the cocoon (dashed), shocked ISM (dotted), and beam (solid) for model f005m1C.
    }
    \label{fig:timeevo1}
\end{figure}

This section presents the temperature evolution.
We investigate the time evolution of the energy of electrons and ions.
We divide the whole system of the jet--ICM interaction into four areas that correspond to different physical conditions, namely the beam, the cocoon, the shocked ICM, and the unperturbed ICM.
Figure \ref{fig:divide} presents an example of such division at $t=20.0$ Myr for model f005m1C.
The beam region is identified by a high bulk flow speed.
We define the threshold $v_{\rm z} > 0.9 v_{\rm z, inj}$ to distinguish the beam from the cocoon.
The growth of instabilities makes it difficult to distinguish between the cocoon and the shocked ICM.
However, we assume that the initial ICM is not magnetized.
Therefore, the toroidal field can be used to trace the cocoon, i.e., $|B_{\rm \phi}| > 0$ for the cocoon.
Finally, the shocked-ICM and unperturbed-ICM regions are distinguished according to whether the gas pressure is higher than its initial value.
Figure \ref{fig:timeevo1} shows the evolution of the volume-weighted density,
\begin{equation}
  \bar{\rho} = \frac{ \iint 2\pi r \rho drdz }{\iint{2\pi r drdz}},
\end{equation}
in the cocoon (dashed), the shocked ISM (dotted), and the beam (solid) for the model f005m1C.
Initially (<0.3 Myr), the backflowing gas directly interacts with the boundary at $z=0$. However, we only consider the period after 0.5 Myr.
The volume-weighted density of the shocked ICM is twice that of the initial ICM density, $20\rho_0$, owing to the shock compression
and does not change with time.
Additionally, the volume-weighted density of the beam mostly remains at its initial level, $0.1\rho_{\rm 0}$.
Meanwhile, the volume-weighted density decreases in the cocoon because of the volume expansion.
A low-density ($\rho \sim 10^{-3} \rho_0$) cavity forms around the hotspots (see Fig. \ref{fig:Fiducial} (a)).
However, the volume of the cavity is small compared with the volume of the cocoon.
The mixing region beside the contact discontinuity has a high density and large radius.
In the mixing region beside the contact discontinuity, the density is low and radius of that is large.
The volume-weighted density of the cocoon thus remains higher than that of the beam gas.

\begin{figure}
  \includegraphics[width=0.95\columnwidth]{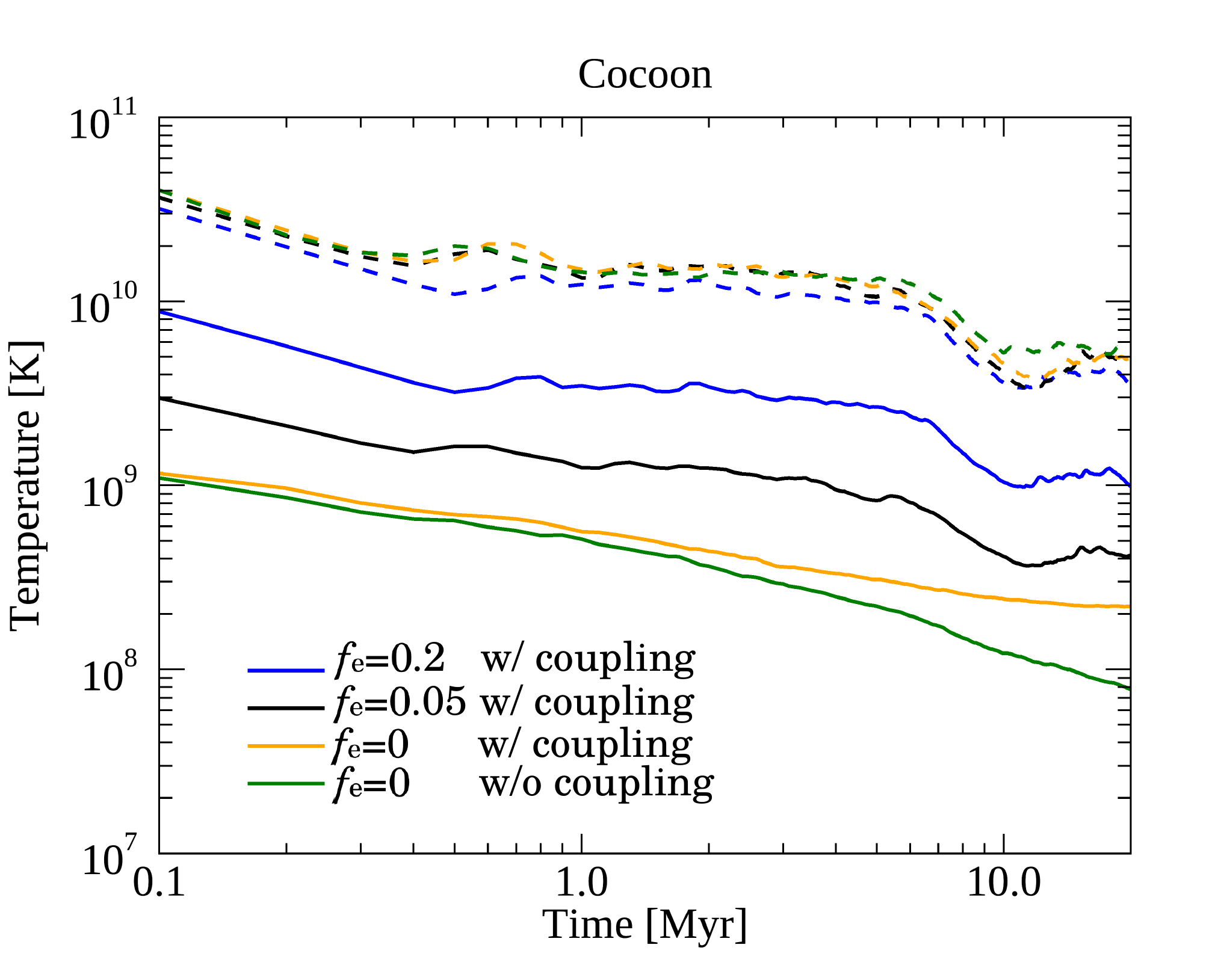}
    \caption{
    Time evolution of the volume-weighted electron (solid) and ion (dashed) temperatures in the cocoon for models f02m1C (blue), f005m1C (black), f00m1C (yellow), and f00m1 (green).
    }
    \label{fig:timeevo2}
\end{figure}

\begin{figure}
  \includegraphics[width=0.95\columnwidth]{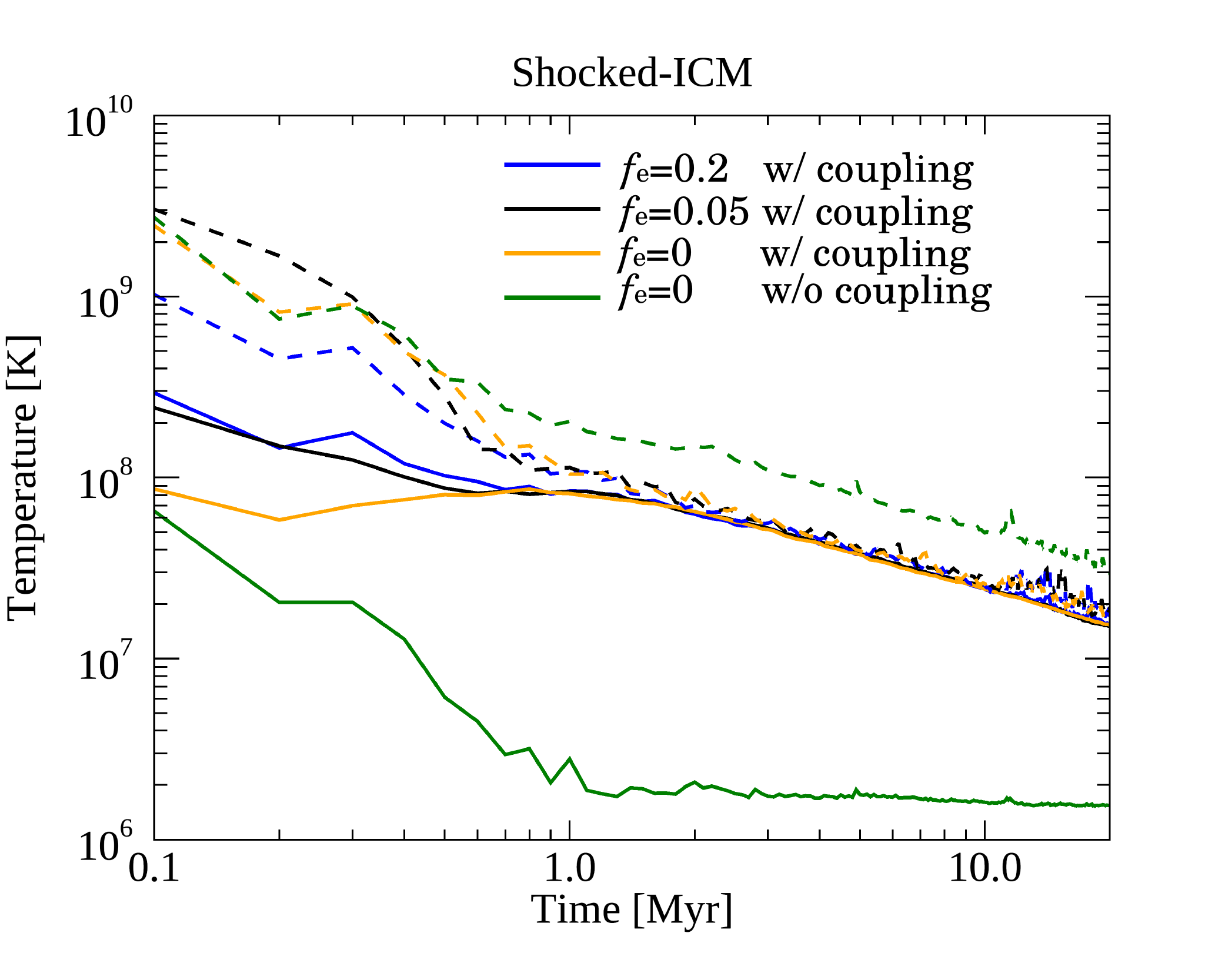}
    \caption{
    Same as Fig. (\ref{fig:timeevo2}) but in the shocked ICM.
    }
    \label{fig:timeevo3}
\end{figure}

Figures \ref{fig:timeevo2} and \ref{fig:timeevo3} show the evolution of the volume-weighted average electron and ion temperatures,
\begin{equation}
  \bar{T}_{\rm i} = \frac{\iint{2\pi r T_{\rm i}} drdz }{\iint{2\pi r drdz}},~~\bar{T}_{\rm e} = \frac{\iint{2\pi r T_{\rm e}} drdz }{\iint{2\pi r drdz}},
\end{equation}
in the cocoon and the shocked ICM as a function of time for f002m1C, f005m1C, f00m1C, and f00m1.
The electron temperature of the cocoon strongly depends on $f_{\rm e}$ because the shocked gas heated at the terminal shock forms the cocoon.
Meanwhile, the ion temperature is not sensitive to $f_{\rm e}$ in the range $f_{\rm e} < 0.2$.
The heating time scale of electrons, $t_{\rm heat} \equiv nk_{\rm B}T_{\rm e}/q^{\rm ie}$, is about $10^{2-3}$ Myr in the cocoon.
Thus, electrons and ions are not in thermal equilibrium in the cocoon.
The volume-weighted temperature in the cocoon decreases because the gas in the mixing region, where $r$ is large, makes a large contribution to the volume-weighted temperature.
Coulomb coupling does not make a large contribution to electron heating in the case of the models with $f_{\rm e}=$ 0.05 and simulation time of 0.2 Myr.
If we carry out longer simulations, we expect that the temperatures become low and Coulomb coupling becomes dominant.
In contrast, electrons warm appreciably through Coulomb coupling in the case that $f_{\rm e} = 0$ (f00m1 and f00m1C in Fig. \ref{fig:timeevo2}).
For all models, ions are hardly affected by Coulomb coupling and, the ion temperature remains high.
Coulomb coupling dominates in the gas mixing region, and the electron temperature remains high even in the case that $f_{\rm e} > 0$ after $t>10$ Myr.

In the shocked ICM, the three models with Coulomb coupling reach thermal equilibrium between ions and electrons within 1--2 Myr.
Coulomb coupling affects electrons and ions regardless of $f_{\rm e}$ because the shocked ICM is denser than the cocoon gas.
We thus see that the heating time scale in the shocked ICM is shorter than 1 Myr.
The model f00m1, which does not have Coulomb coupling, is still in a two-temperature state at the end of the simulation.
The electron temperature decreases but does not fall below the temperature of the initial ICM.

\begin{figure}
  \includegraphics[width=0.95\columnwidth]{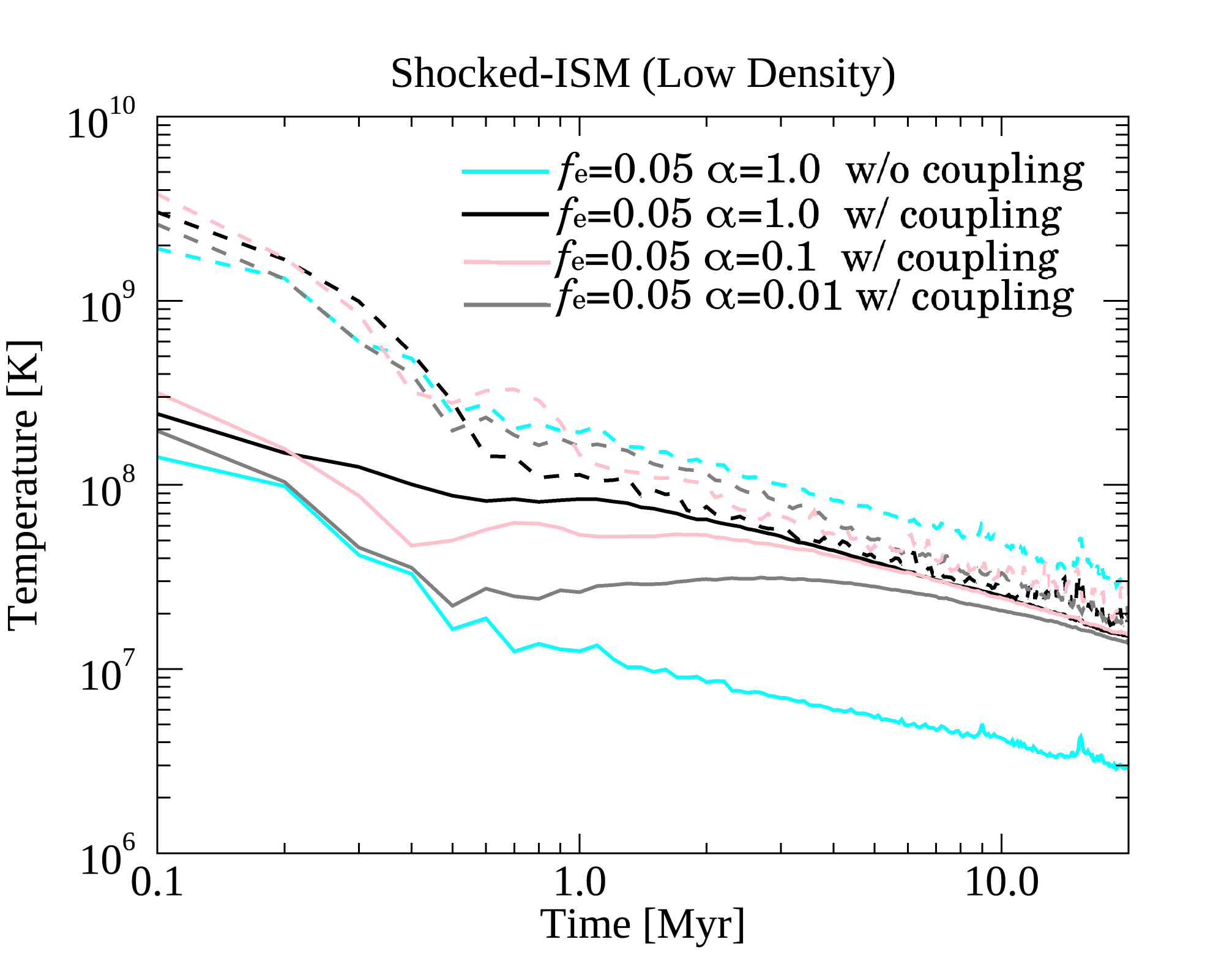}
    \caption{
    Same as Fig. (\ref{fig:timeevo3}) but in the shocked ICM for models f005m1C (black), f005m1C$\alpha$-1 (pink), f005m1C$\alpha$-2 (gray), and f005m1 (cyan).
    }
    \label{fig:timeevo4}
\end{figure}

Figure \ref{fig:timeevo4} is the same as Fig. \ref{fig:timeevo3} but for models f005m1C (black), f005m1C$\alpha$-1 (pink), f00m1C$\alpha$-2 (gray), and f005m1 (cyan) in the shocked-ICM.
Here, f005m1C$\alpha$-1 and f005m1C$\alpha$-2 are respectively the models for which the density parameters are set $10^{-1}$ and $10^{-2}$ times lower than the density for the model f005m1C.
Coulomb coupling weakens as we reduce the normalized density.
Note that the dynamics do not depend on the normalized density because the dynamics of non-relativistic jets are determined by the density ratio of the jet beam to the ICM and the magnetosonic Mach number.
Moreover, Coulomb coupling does not affect the total thermal energy.
We clearly see that the relaxation time between electrons and ions is longer when the density parameter $\alpha$ is small.
We expect that the relaxation times for the models f005m1C$\alpha$-1 and f005m1C$\alpha$-2 are respectively $10^2$ and $100^2$ times the relaxation time for the model f005m1C because Coulomb coupling is proportional to the normalized density.
However, electrons and ions reach an equilibrium within 10 Myr for the model f00m1C$\alpha$-2.
This is because the heating time scale of electrons strongly depends on the electron temperature\footnote{
The ratio of energy transfer through Coulomb coupling is written as $q^{\rm ie}\propto (T_{\rm i}-T_{\rm e})(\sqrt{\pi/2}+\sqrt{\theta_{\rm i}+\theta_{\rm e}})(\theta_{\rm i}+\theta_{\rm e})^{-3/2}$.
We here assume $T_{\rm e}=10^{8}$K and $T_{\rm i}=10^{9}$K and hence $\theta_{\rm e}\sim 10^{-2}$ and $\theta_{\rm i}\sim 10^{-4}$ for the shocked ICM.
The equation can be approximated as $q^{\rm ie}\propto (T_{\rm i}-T_{\rm e})\theta_{\rm e}^{-3/2}$.
Thus, the heating time scale of electrons is easily estimated as $t^{\rm ie}_{\rm heat} = nk_{\rm B}T_{\rm e}/q^{\rm ie}\propto T_{\rm e}\theta_{\rm e}^{3/2} (T_{\rm i}-T_{\rm e})^{-1}$.
}
, and the heating time scale is shortened by decreasing both temperatures in the shocked ICM.

In the beam area, the electron and ion temperatures are decoupled by internal shocks.
The volume-weighted ion temperature is about 5 times that of electrons for the model f02m1C.
We estimate that the heating time scale of electrons is about 10 Myr.
This value is within our simulation time.
In practice, it is difficult to reach thermal equilibrium between electrons and ions because the gas flows into the cocoon continuously within a short time.
Furthermore, ions are primarily heated by the internal shocks.
\section{Discussion}
\label{sec:discussion}
\subsection{Electron Heating at Shock Waves}
\label{sec:fe}
This section discusses an appropriate $f_{\rm e}$ value for AGN jets, focusing on the bow shock and the terminal shock.
The Mach number of the bow shock is greater than 5 because the temperature of the shocked ICM is 10 times the initial temperature, and the Mach number of the terminal shock is about 14.

Observations of bow shocks of the Earth and Saturn and a supernova remnant indicate that the post-shock temperature ratio of ions to electrons is proportional to the magnetosonic Mach number.
\cite{2015A&A...579A..13V} derived the equation for the post-shock ion--electron temperature ratio assuming adiabatic heating of electrons and heat exchange between electrons and ions using the exchange factor $\zeta$, which is defined as the fraction of the enthalpy-flux difference between ions and electrons.
An appropriate value that explains the observational results is $\zeta = 5\%$.
The parameter $\zeta$ corresponds to the fraction of the electron heating $f_{\rm e}$ in our work, assuming thermal equilibrium between electrons and ions in the pre-shock region.

The energy exchange ratio in collisionless shocks strongly depends on the development of microscale instabilities, which is affected by the Mach number, the plasma $\beta$, the pre-shock temperature ratio, the shock angle, and other factors.
Therefore, the theoretical derivation of the fraction of electron heating in collisionless shocks still has large uncertainties.
However, some theoretical studies of collisionless shocks showed that shocks primarily heat ions.
\cite{2018ApJ...858...95G} carried out two-dimensional kinetic particle-in-cell simulations of low-Mach-number shocks, assuming galactic shocks, and showed $T_{\rm e}/T_{\rm i} \sim 0.24$ at ${\mathcal M}=5$, independent of the plasma beta ranging $4 < \beta < 32$.
In addition, \cite{2010PhPl...17d2901M} found that the post-shock temperature ratio is proportional to the magnetosonic Mach number and $T_{\rm e}/T_{\rm i} = 0.01$ when the shock parameters are $\beta=10$ and ${\mathcal M}=14$.

The above results indicate that the models for which $f_{\rm e}=0.2$ overestimate the post-shock electron temperature.
Meanwhile, the model for which $f_{\rm e}=0$ (i.e., electrons are heated only by shock compression) is reasonable for low-Mach-number shocks.
However, this model is not appropriate for high-Mach-number shocks because some instantaneous electron heating mechanism is needed.
It is noted that instantaneous heating occurs on a time scale shorter than the time scale of Coulomb collision.
We therefore argue that the value 0.05 for $f_{\rm e}$ is slightly high but most reasonable in this work.

\subsection{Cooling and Heating Time Scale}
The distribution of radiative intensity and the cooling time scale for two-temperature plasma are different from those for one-temperature plasma.
In this subsection, we consider bremsstrahlung radiation for a thermal distribution of electrons as radiative cooling \citep{1986rpa..book.....R}:
\begin{equation}
  q^{\rm ff} \sim T_{\rm e}^{1/2}  n^2 (1+4.4\times10^{-10}T_{\rm e}).
\end{equation}
In this work, we assumed that the electron temperature does not exceed the ion temperature anywhere except model f03m1.9C in which the injection temperature of electron is higher than that of ion.
Thus, the cooling time scale for two-temperature plasma is much longer than that for one-temperature plasma.
The results also indicate that the bremsstrahlung cooling for one-temperature plasma is much stronger than that for two-temperature plasma.
It is difficult to observe thermal radiation from a cocoon of an AGN jet because the density is low there.
However, \cite{2007MNRAS.376.1630K,2009MNRAS.395L..43K} showed the possibility of a young radio-loud AGN emitting thermal MeV--GeV $\gamma$-ray bremsstrahlung radiation.
The detection limit is sensitive to the thermalization of electrons and ions in the cocoon.

We next estimate the cooling and heating time scale for plasma, ions, and electrons:
\begin{equation}
  t^{\rm ff}_{\rm cool,gas} = nk_{\rm B}T_{\rm gas}/q^{\rm ff}(n^2,T_{\rm e}),
\end{equation}
\begin{equation}
  t^{\rm ie}_{\rm cool,i} = nk_{\rm B}T_{\rm i}/q^{\rm ie}(n^2,T_{\rm e},T_{\rm i}),
\end{equation}
\begin{equation}
  t^{\rm ff/ie}_{\rm cool/heat,e} = nk_{\rm B}T_{\rm e}/(q^{\rm ff}(n^2,T_{\rm e})-q^{\rm ie}(n^2,T_{\rm e},T_{\rm i})).
\end{equation}
Note that the ratio of the cooling and heating time scale does not depend on density because both bremsstrahlung radiation and Coulomb coupling are proportional to the square of density.
In addition, electrons are heated by Coulomb coupling if $q^{\rm ie} > q^{\rm ff}$.
If the ion temperature is much higher than the electron temperature and electrons are in a transrelativistic regime, the energy transfer rate via Coulomb coupling is proportional to the ion temperature, $q^{\rm ie} \propto T_{\rm i} - T_{\rm e} \sim T_{\rm i}$ (see eq.\ref{eq:qie}).
The cooling time scale for ion gas therefore depends on only the gas density, $t^{\rm ie}_{\rm cool,i} \sim 28 n^{-1}$ Myr.

When the ion temperature is higher than the electron temperature, $T_{\rm i} \sim 10T_{\rm e}$, the energy transfer rate $q^{\rm ie}$ for Coulomb coupling is larger than the bremsstrahlung-energy loss rate $q^{\rm ff}$.
This means that the electrons continue to be heated by ions.
The heating time scale of electrons decreases in proportion to the temperature difference between electrons and ions ($\sim$ plasma).
At this time, ions act as a heat bath for electrons because the cooling time scale for ions is much longer than the heating time for electrons.
In particular, when $T_{\rm gas}=10^{10}$ K and $T_{\rm i} > 0.3 T_{\rm e}$, the cooling time scale of ions is 10 times the heating time for electrons.

In this section, we considered only bremsstrahlung radiation and Coulomb coupling.
However, viscous heating may affect the temperature evolution in the cocoon where KH instability is developing. (We discuss viscous heating in the next section.)
Moreover, other cooling processes, namely thermal synchrotron cooling and Compton cooling, may become dominant.
These emission processes are more complicated than the process of bremsstrahlung emission.
Simulations including these processes and viscous heating are therefore the next step of our study.
\subsection{Viscous heating due to turbulence in the cocoon}
Vortex motions develop around the surface between the cocoon and the shocked ICM.
Our simulations, however, have no explicit means of dissipating sound waves because we use ideal MHD equations.
Viscosity induces effective energy diffusion because kinematic viscosity becomes remarkably high in high-temperature plasma, such as the ICM and jets \citep{1965RvPP....1..205B}.
The exact value of the heating rate in MHD wave damping is still under debate.
Previous two-temperature MHD works on accreting flow used a simple fitting formula based on theoretical models of the dissipation of MHD turbulence in weakly collisionless plasmas developed by \cite{2010MNRAS.409L.104H}.
\cite{2019PNAS..116..771K} carried out numerical simulations using a hybrid fluid--gyrokinetic model and updated the results of \cite{2010MNRAS.409L.104H}.
The fitting formula of \cite{2019PNAS..116..771K} (hereafter K19) is
\begin{equation}
  \label{eq:kawazura1}
  \frac{Q_{\rm i}}{Q_{\rm e}} = \frac{35}{1+(\beta_{\rm i}/15)^{-1.4} \exp{(-0.1 T_{\rm e}/T_{\rm i})} },
\end{equation}
\begin{equation}
  \label{eq:kawazura2}
  f_{\rm e,turb} = \frac{1}{1 + Q_{\rm i}/Q_{\rm e} },
\end{equation}
where $Q_{\rm i}, Q_{\rm e}$, and $\beta_{\rm i} \equiv nk_{\rm B}T_{\rm i}/8 \pi B^2$ are respectively the heating rates of ions and electrons and the ratio of ion thermal energy to magnetic energy.
$f_{\rm e,turb}$ is almost insensitive to $T_{\rm i}/T_{\rm e}$ but depends on $\beta_{\rm i}$ strongly.
When thermal energy is dominant ($\beta_{\rm i} > 1$), the turbulence heats primarily ions; i.e., $Q_{\rm i}/Q_{\rm e} > 1$.
In contrast, electrons receive most of the heat at low $\beta_{\rm i}$.

\begin{figure}
	\includegraphics[width=0.95\columnwidth]{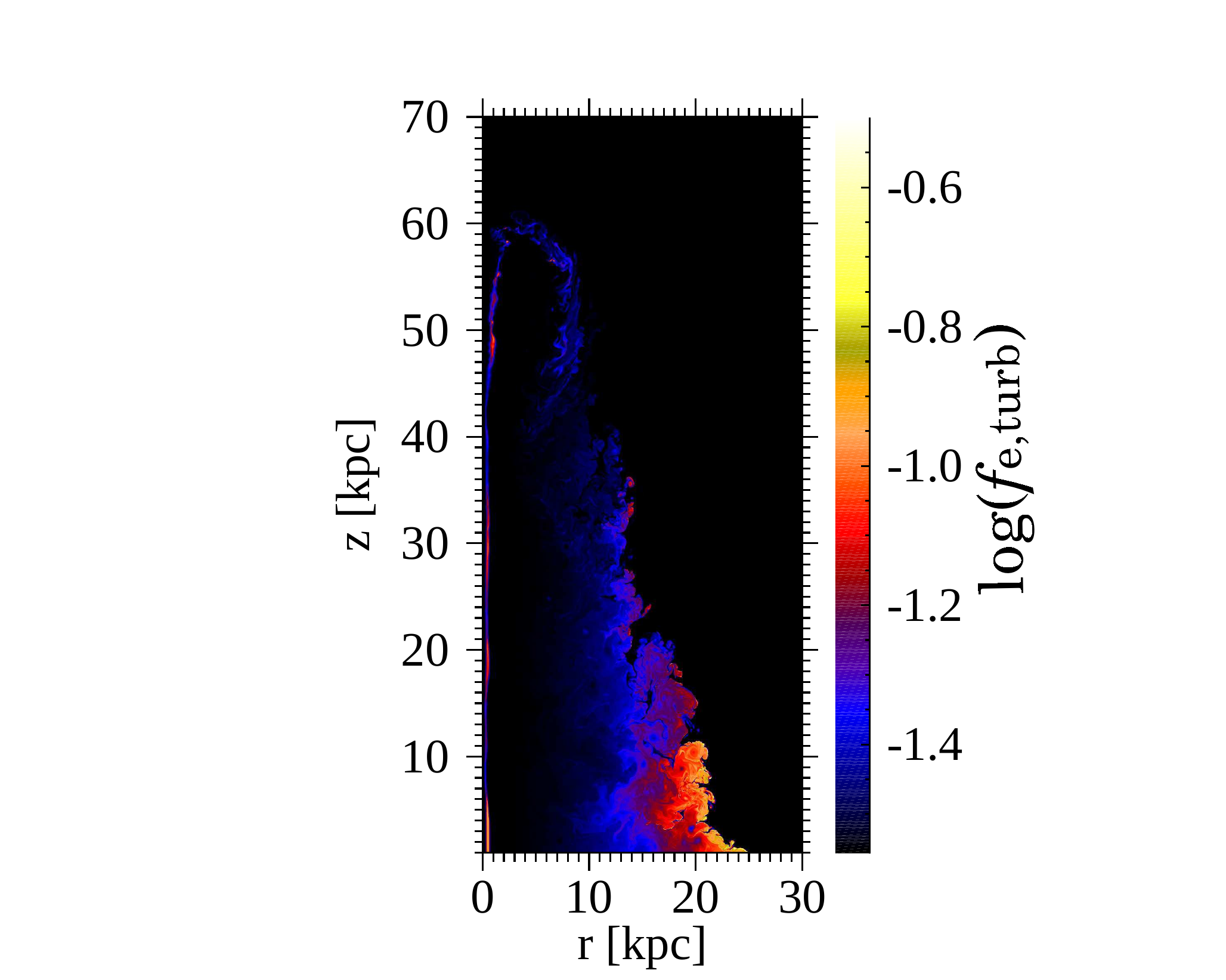}
    \caption{
Snapshot of the fraction of electron heating $f_{\rm e,turb}$ estimated for model f005m1C at $t=20$ Myr using eq. (\ref{eq:kawazura1}) and eq. (\ref{eq:kawazura2}).
    }
    \label{fig:femap}
\end{figure}

To estimate the effect of turbulence heating, we calculate $f_{\rm e,turb}$ from eq. (\ref{eq:kawazura1}) and eq. (\ref{eq:kawazura2}) for model f005m1C at $t=20$ Myr (Fig. \ref{fig:femap}).
Note that K19 is not a suitable model for shocks because the electron heating in shocks is more complicated, which depends on pre-shock physical quantities.
Figure \ref{fig:femap} shows that most of the dissipated energy goes to ions.
While magnetic energy accumulates along the contact discontinuity, the pressure of ions still dominates.
Therefore, $f_{\rm e,turb}$ is lower than 0.2; i.e., turbulent heating is inefficient for electrons.
Of course, if we consider Poynting-flux-dominated jets, electrons could be heated more efficiently than ions.
However, it is expected that most magnetic energy is converted to the kinetic energy of bulk motion at sub--parsec scales.
Moreover, a three-dimensional magnetic kink instability develops, and the magnetic energy of the jet is converted to internal energy rapidly \citep{2015MNRAS.452.1089P}.

\cite{2015ApJ...803...48G} carried out hydrodynamical simulations of the formation and evolution of X-ray cavities in the ICM formed by jets taking into account the kinetic viscosity.
They showed that viscosity affects the shape of cavity and suppresses KH instability between the cocoon and the shocked ICM.
Therefore, the viscosity provides an efficient energy dissipation mechanism.
However, the anisotropic Braginskii viscosity, which transports momentum along the orientation of the magnetic field, cannot suppress KH instability because the toroidal magnetic field is dominated in the cocoon and the effect of viscosity on the velocity shear across the magnetic field lines is inefficient \citep{2013ApJ...768..175S}.
\subsection{Equation of State}
Our simulations use the non-relativistic ideal equation of state for both electrons and ions.
However, the specific heat ratio changes from 5/3 for a non--relativistic (cold) plasma to 4/3 for a relativistic (hot) plasma.
In particular, the relativistic temperature of electrons is about $3.0\times10^9$ K.
Electrons in the jet have therefore reached their relativistic temperature, and the specific heat ratio of electrons becomes 4/3.
Meanwhile, ions are non-relativistic in our simulation because the relativistic temperature of ions is about $10^{13}$ K.
The effective adiabatic index for the gas can be calculated as \citep{2015MNRAS.454.1848R}
\begin{equation}
  \gamma_{\rm gas}-1=\frac{p_{\rm i}+p_{\rm e}}{\epsilon_{\rm i}+\epsilon_{\rm e}}=(\gamma_{\rm e}-1)(\gamma_{\rm i}-1)\frac{1+T_{\rm i}/T_{\rm e}}{(\gamma_{\rm i}-1)+(\gamma_{\rm e}-1)T_{\rm i}/T_{\rm e}}.
  \label{eq:gm}
\end{equation}
Thus, $\gamma_{\rm gas}$ becomes 13/9 if electrons are relativistic.

A small value of the specific heat ratio leads to large internal energy.
Therefore, a high value of internal energy accelerates the propagation to increase momentum flux in the relativistic regime, $p_{\rm gas} \sim \rho c^2$ \citep{2007MNRAS.378.1118M}.
In the non-relativistic regime, however, the momentum flux is almost unaffected by an increase in internal energy.
The variation in the specific heat ratio is therefore negligible in terms of jet dynamics in our simulations.
Of course, we note that a softer equation of state leads to a lower sound speed and increases the interval of internal shocks.
If the knots of AGN jets correspond to internal shocks, then the positions of the knots depend on the specific heat ratio.

A variation in the specific heat ratio is negligible in terms of dynamics, but the low specific heat ratio may decrease the electron temperature.
The first reason is that the temperature jump condition depends on the specific heat ratio (see eq. \ref{eq:rankine-hugonion}).
The second reason is that more energy is required to heat electrons through Coulomb coupling  when the specific heat ratio is small.
Therefore, electrons and ions will more readily have different temperatures when we adopt the relativistic equation of state for electrons.
\subsection{Observational Implications}
In radio galaxies such as Cygnus A, X-ray cavities are observed in the radio lobe around the jet \citep{2006ApJ...644L...9W}.
These cavities can be explained by the low density, hot plasma in the cocoon (see Figure 1(a)).
Meanwhile, X-ray emission is enhanced in the shocked ICM surrounding the cocoon (see fig. \ref{fig:Fiducial}(h) and fig. \ref{fig:divide}).
Diffuse soft X-rays observed around the core of the AGN can be emitted from the cocoon plasma mixed with the ICM  through Kelvin-Helmholtz instability.
Our numerical results (Fig. \ref{fig:Fiducial}(h)) indicate that the electron temperature in this region is around $T_{\rm e}=10^8-10^9$K.
This region can therefore enhance X-ray emission.

Observations of radio lobes indicate that the cocoon expands as the cocoon plasma flows back toward the galactic center.
Therefore, the cocoon pressure should be higher than the ICM pressure.
In a single temperature plasma, it is not consistent with the observations of both FRI/II jets which show that the electron pressure in the lobe is lower than the
external pressure, i.e., the lobe is under-pressured \citep[e.g.,][]{2007MNRAS.381.1109B}.
Our numerical results resolve this problem because the ion pressure much exceeds the electron pressure in sub-relativistic AGN jets.

Electrons are in a relativistic regime in our simulations.
Non-thermal particle acceleration for electrons is therefore efficient in both Fermi/diffusive shock acceleration and turbulent dissipation \citep{2019PhRvL.122e5101Z,2011ApJ...726...75S}.
The lobes are most prominently observed by the synchrotron emission of non-thermal electrons.
It is thus not possible to directly obtain the thermal temperatures of both electrons and ions in radio observations.
However, the efficiency of particle acceleration theoretically depends on the thermal energy of electrons and ions.
Therefore, the population of non-thermal electrons depends on that of the thermal electrons.
Groups have developed post-processing code for calculating non-thermal electron spectra from the results of MHD simulations \citep{2018ApJ...865..144V,2019MNRAS.488.2235W}.
In order to obtain the realistic non-thermal electron spectra of AGN jets, it is necessary it use the thermal electron temperature  obtained from two-temperature MHD simulations.

Diffuse thermal X-ray emissions have been detected in the radio lobes of a few sources, such as Fornax A and Centaurus A \citep{2013PASJ...65..106S,2013ApJ...766...48S}.
The thermal emission comes from the mixing region between the radio lobe and the shocked ICM.
In our work, two temperature plasma still exists in the mixing region (see Fig. \ref{fig:Fiducial}(i) $z<20$ kpc, $10<r<20$ kpc).
Since the thermal Doppler broadening of spectral lines gives information about each the ions thermal velocity, future X-ray observations with high-resolution spectroscopy(e.g., the X-ray Imaging and Spectroscopy Mission, XRISM) will reveal the presence of the two-temperature plasma in cocoons of AGN jets.

\section{Summary and Conclusions}
\label{sec:summary}
We performed two-dimensional and axisymmetric simulations of AGN jet propagation into a constant-density ICM to study the fraction of electron heating, which affects the electron temperature distribution.
\begin{itemize}
  \item  In axisymmetric jets, the energies of electrons and ions are decoupled at internal shocks.
         In particular, ions have about twice the internal energy of electrons in the downstream of the first internal shock.
         As in the one-dimensional case, the temperature ratio is described by eq.(\ref{eq:post-shock temp}) in the downstream of the terminal shock.
         However, post-shock electrons lose energy through adiabatic expansion in the multidimensional case
         and the temperature ratio may therefore fall below the value predicted using eq. (\ref{eq:post-shock temp}) at low values of $f_{\rm e}$.
         Furthermore, we found that the temperature of the injected jet does not affect the temperature in terminal regions when $f_{\rm e}$ is constant.
  \item The volume-weighted temperature of the cocoon decreases as the region expands.
        Coulomb coupling is weak in the cocoon because the electrons have reached relativistic temperatures, and the density is low.
        The time scale of energy transfer due to Coulomb collision is therefore about $10^{2-3}$ Myr at the end of simulations, and the electrons continue to be heated by ions in the cocoon.
        These results indicate the existence of two-temperature plasma in the X-ray cavity.
        In the shocked ICM, the ion thermal energy is converted to electron thermal energy efficiently, and ions and electrons achieve thermal equilibrium in 1 Myr.
        Moreover, mixing of the jet plasma and shocked ICM through KH instability at the interface could enhance soft X-ray emissions around the contact discontinuity between the cocoon and shocked ICM.
  \item We investigated the density dependence of the volume-weighted temperature evolution in the shocked ICM.
        The time scale of relaxation between electrons and ions is certainly extended for a lower-density model.
        However, the electron heating time scale strongly depends on the electron temperature, with a lower electron temperature resulting in a shorter heating time scale.
        Therefore, the lowest-density model, which has a density 100 times lower than that of the fiducial model, achieves thermal equilibrium by 10 Myr.
\end{itemize}

Our work indicates that two-temperature plasma exists in the X-ray cavity, and electrons and ions probably reach thermal equilibrium in the shocked ICM.
However, the determination of the ion temperature is a challenge.
XRISM observations could provide useful information on ion thermal energies from a line profile.

\section*{Acknowledgements}
We are grateful to Dr. Y. Matsumoto and Dr. S. Matsukiyo for helpful discussions.
We thank the anonymous referee for helpful suggestions.
Our numerical computations were carried out on the Cray XC50 at the Center for Computational Astrophysics of the National Astronomical Observatory of Japan and the Fujitsu PRIMERGY CX600M1/CX1640M1 (Oakforest-PACS) at the Information Technology Center, The University of Tokyo.
This research is partially supported by an Initiative on Promotion of Supercomputing for Young or Women Researchers, Information Technology Center, The University of Tokyo.
This work is supported by JSPS KAKENHI Grant Numbers 19K03916 and 16H03954.
We thank Glenn Pennycook, MSc, from Edanz Group (www.edanzediting.com/ac) for editing a draft of this manuscript.




\bibliographystyle{mnras}
\bibliography{mnras2} 



\appendix
\section{One-dimensional Jet Simulations }
\subsection{Numerical Setup}
We assume two separate zones in simulating the propagation of jets.
The left side is a jet beam having a low number density ($n=0.005$/cc), high temperature ($T_{\rm i}=T_{\rm e}=10^9$ K), and bulk velocity of 0.2c.
The right side is assumed to be an ICM having high density ($n=0.5$/cc) and low temperature ($T_{\rm i}=T_{\rm e}=10^6$ K).
We make calculations for four models with different fractions of electron heating $f_{\rm e} = 0, 0.05, 0.2, 0.5$.
In the one-dimensional simulation, we neglect the magnetic field and energy exchange through Coulomb coupling.
The computational domain is $x/r_{\rm jet} \in [0,10]$ and the number of grid points is 1024.
\subsection{Results of One-dimensional Simulations}
\label{sec:1d}
\begin{figure}
	\includegraphics[width=0.95\columnwidth]{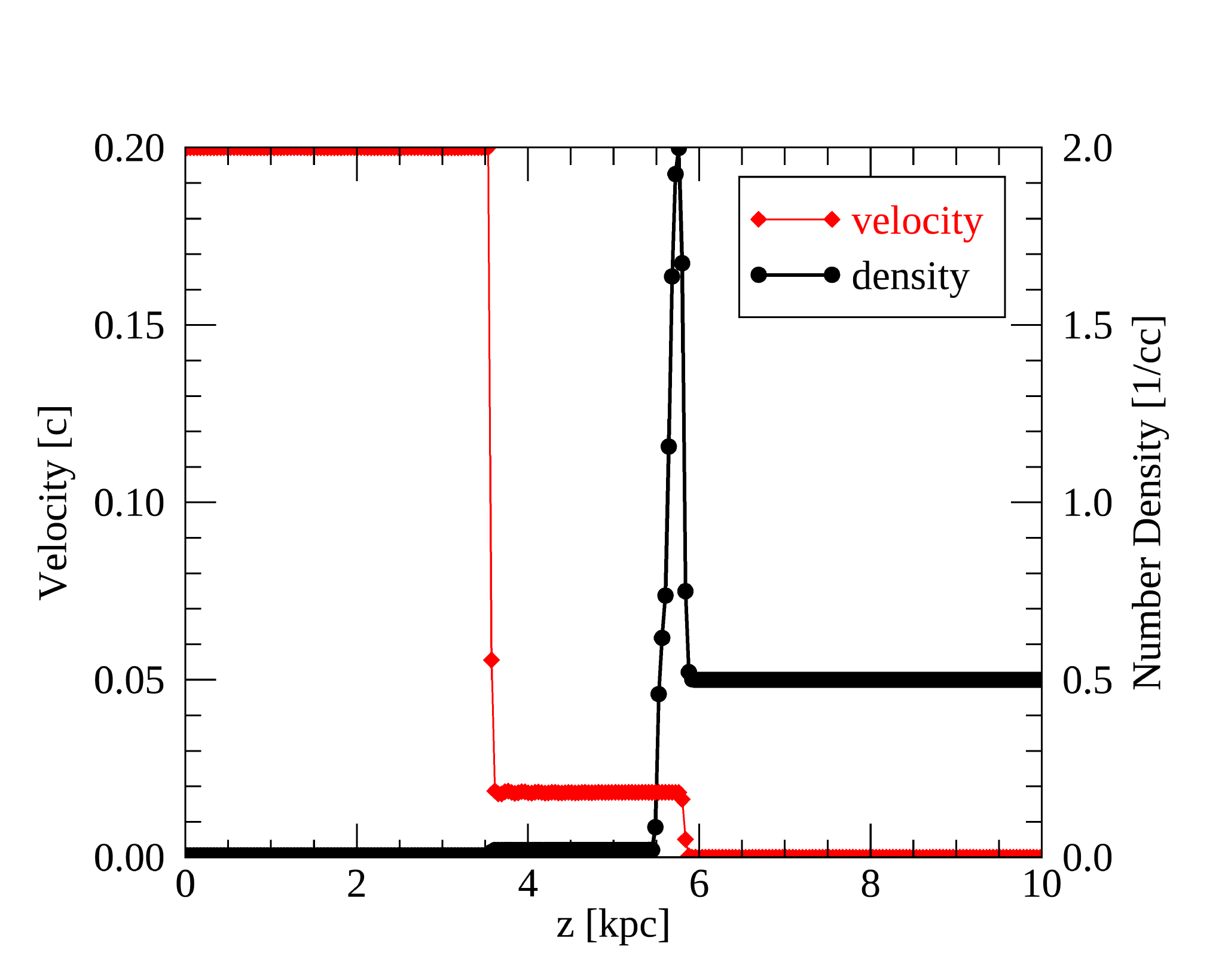}
    \caption{Velocity (red) and density (black) profiles for one-dimensional simulation of the propagation of a supersonic jet at $t=0.1$ Myr.
    We capture the reverse shock (i.e., terminal shock), contact discontinuity, and forward shock (i.e., bow shock).
    }
    \label{fig:1djetrov}
\end{figure}

\begin{figure}
	\includegraphics[width=0.95\columnwidth]{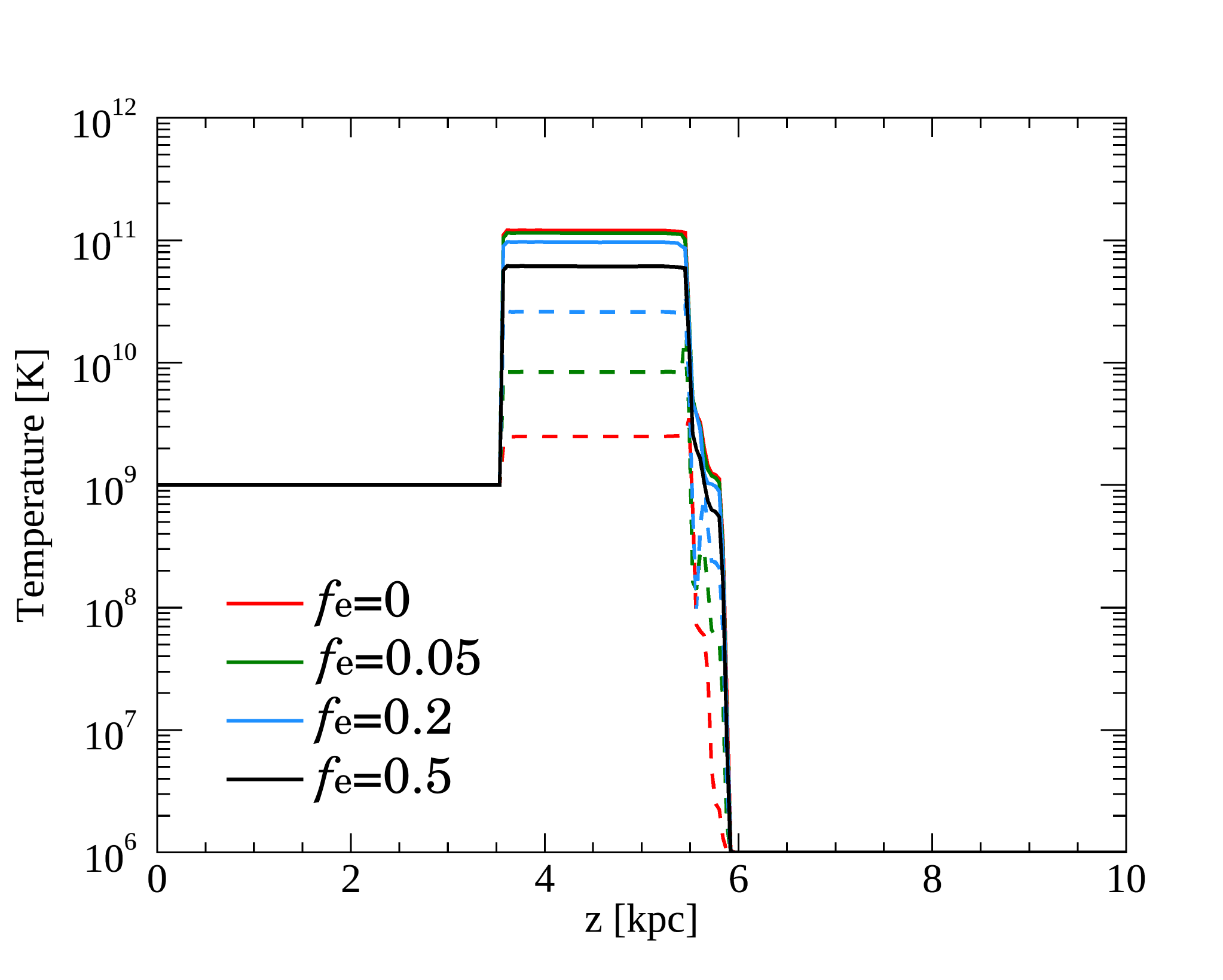}
    \caption{
    Same as Fig. \ref{fig:1djetrov} but showing ion (solid) and electron (dashed) temperatures.
    The line colors represent the fraction of electron heating (red: $f_{\rm e}=0$, green: $f_{\rm e}=0.05$, blue: $f_{\rm e}=0.2$, black: $f_{\rm e}=0.5$).
    Ions and electrons heat up at both shocks,
    and the fraction of electron heating affects the electron temperature appreciably.
    }
    \label{fig:1djetteti1}
\end{figure}

Figure \ref{fig:1djetrov} shows the density (black) and velocity (red) profiles at $t=0.1$ Myr.
The forward shock (bow shock), contact discontinuity, and reverse shock (terminal shock) are easily identified.
Because the bow shock compresses the ICM, a high-density shocked ICM forms between the contact discontinuity and bow shock.
Figure \ref{fig:1djetteti1} shows the temperature distribution at the same time as for the results in Fig. \ref{fig:1djetrov}.
Solid and dashed lines respectively denote ion and electron temperatures.
Colors represent fractions of electron heating of $f_{\rm e} =$ 0.0 (red), 0.05 (green), 0.2 (blue), and 0.5 (black).
Because most kinetic energy of the jet dissipates around the reverse shock, a high-temperature region called a hotspot forms between the contact discontinuity and reverse shock.
The gas temperature of the hotspot is obtained by applying the Rankine--Hugoniot jump condition at the reverse shock:
\begin{equation}
  \label{eq:rankine-hugonion}
  \frac{T_{\rm post}}{T_{\rm pre}} = 1+\frac{ 2(\gamma_{\rm gas}-1)(\gamma_{\rm gas}M_{\rm pre}^2+1)(M_{\rm pre}^2-1) }{(\gamma_{\rm gas}+1)^2 M_{\rm pre}^2},
\end{equation}
where $T_{\rm pre} = 10^9$ K is the pre-shock temperature, $T_{\rm post}$ is the post-shock temperature, and $M = 14$ is the Mach number.
We thus derive the hotspot temperature $T_{\rm post} = 6.2\times10^{10}$ K.
In the case that $f_{\rm e} = 0.5$, the dissipative energy is divided to electrons and ions equally.
The post-shock temperatures of the gas, electron, and ion are thus equal ($T_{\rm gas,pos} = T_{\rm e,pos} = T_{\rm i,pos}$) because of the same initial temperatures.
Numerical values of the post-shock temperatures of ions and electrons are $6.14\times10^{9}$ K and are in good agreement with the theoretical values.

In the case that $f_{\rm e} = 0$, electrons purely evolve adiabatically; i.e., the entropy of electrons is conserved through shocks.
The post-shock electron temperature is thus expressed as
\begin{equation}
  \label{eq:fe0}
  \frac{T_{\rm e,post}}{T_{\rm e,pre}} = \eta^{\gamma-1} = 4^{2/3} \sim 2.5 T_{\rm e,post} = 2.5 \times 10^{9}{\rm K}.
\end{equation}
Here, $\eta$ is the shock compression ratio, whose value is 4 at a strong shock when the specific heat ratio is 5/3.
Meanwhile, when $f_{\rm e}$ does not equal zero, electrons receive dissipative energy from the shock.
We therefore obtain the post-shock electron temperature as
\begin{equation}
  \label{eq:post-shock temp}
  T_{\rm e,post} = \eta^{\gamma-1}T_{\rm e,pre} + 2.0 f_{\rm e} ( T_{\rm gas,post} - \eta^{\gamma-1} T_{\rm gas,pre} ).
\end{equation}
The post-shock temperature ratio of the electron to ion predicted using eq. (\ref{eq:post-shock temp}) is 0.073 and 0.269 at $f_{\rm e} = $0.05 and 0.2, respectively.
The post-shock temperature ratio of the electron to ion is actually $0.0735$ and $0.269$ at $f_{\rm e} = $0.05 and 0.2, respectively, in our simulation.
%
\bsp	
\label{lastpage}
\end{document}